\documentclass[aps,prb,twocolumn,superscriptaddress,longbibliography]{revtex4-2}

\usepackage{amssymb,amsmath,amsthm,bbm,bbold}
\usepackage{amsfonts}
\usepackage{graphicx}
\usepackage{mathtools}
\usepackage{hyperref}

\usepackage[normalem]{ulem}
\usepackage{color}
\usepackage{xcolor}

\def\RR{\mathbbm{R}}

\newcommand{\F}{\mathcal{F}}
\newcommand{\bd}[1]{\boldsymbol{#1}}
\newcommand{\wb}{\bd{w}}

\newcommand{\gh}{\hat{\gamma}}
\newcommand{\Gh}{\hat{\Gamma}}
\newcommand{\gwb}{\gh_{\wb}}
\newcommand{\Gwb}{\Gh_{\wb}}

\newcommand*\xbar[1]{\hbox{\vbox{
       \hrule height 0.6pt 
       \kern0.3ex
       \hbox{%
         \kern-0.2em
         \ensuremath{#1}%
         \kern 0.0em
         }}}}

\newcommand*\xxbar[1]{\hbox{\vbox{
       \hrule height 0.6pt 
       \kern0.3ex
       \hbox{%
         \kern-0.0em
         \ensuremath{#1}%
         \kern 0.0em
         }}}}
\usepackage{accents}

\newcommand{\Fw}{\mathcal{F}_{\!\wb}}
\newcommand{\Fbw}{\,\xbar{\mathcal{F}}_{\!\wb}}

\newcommand{\Gbw}{\,\xbar{\mathcal{G}}_{\!\wb}}

\newcommand{\Ew}{\mathcal{E}^N\!(\wb)}
\newcommand{\ew}{\mathcal{E}^1_N(\wb)}

\newcommand{\Ebw}{\,\xxbar{\mathcal{E}}^N\!(\wb)}
\newcommand{\ebw}{\,\xxbar{\mathcal{E}}^1_N(\wb)}

\newcommand{\ebwS}{\,\scriptsize{\xbar{\mathcal{E}}}\normalsize^1_N\hspace{-0.3mm}(\wb)}
\newcommand{\EbwS}{\,\scriptsize{\xbar{\mathcal{E}}}\normalsize^N\hspace{-0.7mm}(\wb)}

\newcommand{\Tr}{\mathrm{Tr}}

\newcommand{\bra}[1]{\mbox{$\langle #1 |$}}
\newcommand{\ket}[1]{\mbox{$| #1 \rangle$}}

\begin{document}
\title{An exact one-particle theory of bosonic excitations: From a generalized Hohenberg-Kohn theorem to convexified N-representability}
\author{Julia Liebert}
\affiliation{Department of Physics, Arnold Sommerfeld Center for Theoretical Physics,
Ludwig-Maximilians-Universit\"at M\"unchen, Theresienstrasse 37, 80333 M\" unchen, Germany}
\affiliation{Munich Center for Quantum Science and Technology (MCQST), Schellingstrasse 4, 80799 M\"unchen, Germany}

\author{Christian Schilling}
\email{c.schilling@physik.uni-muenchen.de}
\affiliation{Department of Physics, Arnold Sommerfeld Center for Theoretical Physics,
Ludwig-Maximilians-Universit\"at M\"unchen, Theresienstrasse 37, 80333 M\" unchen, Germany}
\affiliation{Munich Center for Quantum Science and Technology (MCQST), Schellingstrasse 4, 80799 M\"unchen, Germany}

\date{\today}

\begin{abstract}
Motivated by the Penrose-Onsager criterion for Bose-Einstein condensation we propose a functional theory for targeting low-lying excitation energies of bosonic quantum systems through the one-particle picture. For this, we employ an extension of the Rayleigh-Ritz variational principle to ensemble states with spectrum $\bd w$ and prove a corresponding generalization of the Hohenberg-Kohn theorem: The underlying one-particle reduced density matrix determines all properties of systems of $N$ identical particles in their $\wb$-ensemble states.
Then, to circumvent the $v$-representability problem common to functional theories, and to deal with energetic degeneracies, we resort to the Levy-Lieb constrained search formalism in combination with an exact convex relaxation. The corresponding bosonic one-body $\bd w$-ensemble $N$-representability problem is solved comprehensively. Remarkably, this reveals a complete hierarchy of bosonic exclusion principle constraints in conceptual analogy to Pauli's exclusion principle for fermions and recently discovered generalizations thereof.
\end{abstract}

\maketitle

\section{Introduction}\label{sec:intro}

A comprehensive understanding of quantum matter requires both the knowledge of the respective ground state properties and the accurate description of excitations. In the case of bosonic quantum systems, particularly prominent examples are collective phenomena in ultracold Bose gases such as excitations in Bose-Einstein condensates (BECs) \cite{Ketterle1996, Cornell1996, Ketterle2000, Davidson2002, Navon2021}, superfluids \cite{Benedek1996, Pohl2013, Esslinger2004} or quantum magnetism \cite{Demler2006, Zinner2014, Schollwoeck2014, Batista2014}. Although wave function based methods allow for an exact treatment, at least in principle, they are only practically feasible for relatively small system sizes due to the exponential growth of the underlying Hilbert space. Conversely, Gross-Pitaevskii theory \cite{pita} and methods based on mean-field approximations in general fail to describe strong correlation effects.
To this end, accurate approaches to strongly correlated many-particle systems with an affordable computational cost are of particular demand.

To motivate and put forth such an approach we recall the Penrose and Onsager criterion \cite{Penrose1956}: a system exhibits BEC whenever the maximum eigenvalue of the one-particle reduced density matrix (1RDM) is proportional to the total particle number. From a conceptual point of view, this clearly identifies one-particle reduced density matrix functional theory (RDMFT) as a promising novel approach to BEC. Actually, its scope would even extend to arbitrary strongly correlated bosonic quantum systems since the 1RDM provides direct insights into the correlation strength through its degree of mixedness.
Despite all these points, the foundation for a bosonic ground state RDMFT was proposed only very recently \cite{Benavides20, LS21}. This is even more remarkable, since its fermionic counter part is enjoying an intensive ongoing development \cite{WK15, BCG15, PG16,SKB17, BCEG17, Schilling2018, Schilling2019, GR19, Buchholz2019, Gritsenko2019-2, Schmidt19, Cioslowski19, Piris19, Mitxelena18, Cioslowski20-1, Cioslowski20-2, Cioslowski20-3, Giesbertz20, Gibney21, Schmidt21, Piris21-1}.

In order to describe many-body quantum systems \emph{comprehensively}, an extension of ground RDMFT to excited states is imperative.
It is the main goal of this work to propose exactly such an RDMFT for calculating bosonic excitations energies. For this, we resort in Sec.~\ref{sec:wRDMFT} to a generalization of the Rayleigh-Ritz variational principle to $\wb$-ensemble states. By proving in this framework a generalization of the  Hohenberg-Kohn \cite{HK} and Gilbert theorem \cite{Gilbert}, we confirm the existence of a corresponding universal functional. To turn this abstract result into a more practical one, we then extend in Sec.~\ref{sec:wRDMFT-LL} the formalism by resorting to the Levy-Lieb constraint search in combination with an exact convex relaxation scheme. This then yields a variational expression for the universal functional. To solve the underlying one-body $N$-representability problem we adapt in Sec.~\ref{sec:wRDMFT-LL} a general methodology that has recently been used for solving the corresponding fermionic problem \cite{Schilling21, LCLS21}. This in turn reveals a remarkable hierarchy of bosonic exclusion principles which we derive and discuss in Sec.~\ref{sec:examples}. Finally, we comment on the implications of the bosonic exclusion principles for bosonic lattice DFT in Sec.~\ref{sec:DFT}.

\section{RDMFT for excited states}\label{sec:wRDMFT}
\subsection{Motivation of functional theories}\label{sec:FTmotiv}
To first motivate RDMFT from a general perspective for both bosons and fermions, we observe that different subfields of the quantum sciences are typically characterized by a \emph{fixed} interaction $\hat{W}$. For instance, in quantum chemistry one considers Coulomb interaction, in condensed matter physics the Hubbard on-site interaction and in the field of ultracold gases the contact interaction. Accordingly, each scientific field refers to a corresponding set of Hamiltonians $\hat{H}$ of interest which is parameterized  by the one-particle Hamiltonian $\hat{h}$ according to
\begin{equation}\label{H}
\mbox{RDMFT:}\quad \hat{H}(\hat{h})\equiv \hat{h}+\hat{W}\,.
\end{equation}
Here, $\hat{h}=\hat{t}+\hat{v}$ consists of both a kinetic energy $\hat{t}$ and an external potential $\hat{v}$, and a possible coupling constant in front of $\hat{W}$ would be absorbed into $\hat{h}$ through a rescaling of the energy. Effectively, the motivation behind RDMFT is to solve, at least in principle, the ground state problem of $\hat{H}(\hat{h})$ for the full class $\{\hat{h}\}$ of one-particle Hamiltonians $\hat{h}$. From a heuristic point of view, this identifies the 1RDM $\gh$ as the natural variable since it is conjugate to the one-particle Hamiltonian $\hat{h}$ according to the Riesz representation theorem. Fixing in addition also the kinetic energy $\hat{t}$ would reduce the space of Hamiltonians of interest to an affine subspace, 
\begin{equation}\label{HDFT}
\mbox{DFT:}\quad \hat{H}(\hat{v})\equiv \hat{v}+\hat{t}+\hat{W}\,, 
\end{equation}
which is then parameterized by the external potential only. Since the corresponding natural variable would be the simpler particle density this reasoning would motivate density functional theory (DFT). In our work, however, we consider a fully variable one-particle Hamiltonian due to the following reasons. First, as it has been noted already in Gilbert's seminal work on ground state RDMFT, absorbing inactive core orbitals in molecular systems leads to additional non-local terms in the one-particle Hamiltonian \cite{Gilbert}. Second, recent advances in the field of ultracold gases make possible the variation of the kinetic energy operator \cite{Guenther13, Schempp15}. Third, by involving the full 1RDM, RDMFT is better suited than DFT for dealing with strongly correlation systems. For instance, according to the \emph{nonfreeness} \cite{Gottlieb05, Gottlieb07, Gottlieb14, Gottlieb17} for pure states $\Gh$,
\begin{eqnarray}\label{nonfreeness}
\mathcal{N}(\Gh) &\equiv& \min_{\Gh'\in \mathcal{M}_{\mathrm{free}}} S(\Gh|| \Gh') \nonumber \\
&=& \begin{cases}
 S(\gh) + S(1-\gh)\,,\quad \mbox{for fermions}\\
 S(\gh) - S(1+\gh)\,,\quad \mbox{for bosons}
\end{cases}
\end{eqnarray}
the 1RDM $\gh$ quantifies in terms of the quantum relative entropy $S(\Gh|| \Gh')  \equiv \mbox{Tr}\big[\Gh\big(\ln(\Gh)-\ln(\Gh')\big)\big]$ the ``distance'' of a quantum state $\Gh$ to the manifold $\mathcal{M}_{\mathrm{free}}$ of free states \cite{Gottlieb05, Gottlieb07, Gottlieb14, Gottlieb17, Turner17, Pachos18, Papic18}. The latter include for fermions (in their closure) the Slater determinants, $f_{\varphi_1}^\dagger \ldots f_{\varphi_N}^\dagger \ket{0}$, and $S(\gh)\equiv-\mbox{Tr}_1[\gh(\ln(\gh)]$ denotes in \eqref{nonfreeness} the von Neumann entropy.

\subsection{Generalized variational principle}\label{sec:variatprin}
The heuristic argument about (affine) spaces of Hamiltonians of interest and emerging  natural variables is turned into a concise statement through the Hohenberg-Kohn theorem and generalizations thereof. To explain this, we focus directly on RDMFT for exited states, also since it contains ground state DFT and RDMFT as special cases. Similar to the latter two also our $\wb$-ensemble RDMFT for excited states will be based on a suitable variational principle. In 1988, Gross, Oliviera, and Kohn proposed the following generalization of the Rayleigh-Ritz variational principle in order to extend DFT to excited states \cite{Gross88_1, Gross88_2, Oliveira88}: Let $\hat{H}$ denote a Hamiltonian on a $D$-dimensional Hilbert space $\mathcal{H}$ with increasingly ordered eigenenergies $E_1\leq E_2\leq \ldots \leq E_D$ and corresponding eigenstates $\ket{\Psi_j}$. Furthermore, let $\boldsymbol{w}\in \RR^D$ be a vector with decreasingly ordered entries $w_1\geq w_2\geq \ldots\geq w_D\geq 0$, and normalization $\sum_{i=1}^D w_i=1$.
$\mathcal{E}(\boldsymbol{w})$ shall denote the set of all density operators with fixed decreasingly ordered spectrum $\mathrm{spec}^\downarrow(\hat{\Gamma}) = \boldsymbol{w}$. Then, the weighted sum of the eigenvalues $E_j$ follows as \cite{Gross88_1, Gross88_2, Oliveira88}
\begin{equation}\label{GOK}
E_{\boldsymbol{w}} \equiv \sum_{j=1}^Dw_j E_j = \min_{\hat{\Gamma}\in \mathcal{E}(\boldsymbol{w)}}\Tr\left[\hat{\Gamma}\hat{H}\right]\,,
\end{equation}
where the minimizer is given by
\begin{equation}\label{Gamma_minimizer}
\hat{\Gamma}_{\boldsymbol{w}} = \sum_{j=1}^Dw_j \ket{\Psi_j}\!\bra{\Psi_j}\,.
\end{equation}
This variational principle shall now be applied to the class \eqref{H} of Hamiltonians of interest. Accordingly, each energy eigenvalue $E_j\equiv E_j(\hat{h})$ becomes a function of the variable one-particle Hamiltonian $\hat{h}$. The same holds for the corresponding eigenstates $\ket{\Psi_j(\hat{h})}$ as long as they are unambiguously defined. This is the case whenever the energies $E_j(\hat{h})$ with non-vanishing weights $w_j$ are non-degenerate. Since realistic electronic systems exhibit often symmetries a few comments are in order here: In applications of $\wb$-ensemble RDMFT and the variational principle \eqref{GOK} in general one restricts the sum on the right side in \eqref{GOK} to just a few ($r$ many) weights, i.e., one considers $\wb$ with $w_1 \geq \ldots \geq w_r > w_{r+1} = \ldots = w_D =0$. Having access to the averaged energy function $E_{\boldsymbol{w}}$ would then allow one to extract the energies of the $r$ energetically lowest eigenstates of $\hat{H}(\hat{h})$, e.g., by taking the derivative with respect to $w_j$ or subtracting averages $E_{\boldsymbol{w}}$ for different choices $\wb$ from each other. 
Moreover, the GOK variational principle and the entire formalism presented in the following can be applied for Hamiltonians with obvious symmetries (giving rise to some good quantum number $Q$),
\begin{equation}
\hat{H}(\hat{h}) \equiv \bigoplus_Q  \hat{H}_Q(\hat{h})  \equiv \bigoplus_Q \left(\hat{h}_Q + \hat{W}_Q\right)\,,
\end{equation}
to each symmetry sector $\mathcal{H}_Q$ separately. For instance, in the context of quantum chemistry, this would mean to exploit the spin-symmetries and simplify $\wb$-RDMFT by dealing with the different spin sectors $Q\equiv S$ or $Q\equiv (S,M)$ separately. Initial approaches contributing to this direction were established in ground state RDMFT in Refs.~\cite{Lathiotakis05, Piris09, RP11, G18, Schilling2019, Piris19, Giesbertz20, Cioslowski20-3, P21}. Accordingly, the most relevant applications of $\wb$-ensemble RDMFT from a practical point of view refer to $r\leq 3$, where $r=1$ corresponds to ground RDMFT.

\subsection{Generalized Hohenberg-Kohn theorem}\label{sec:HK}
In order to prove a generalization of the Hohenberg-Kohn \cite{HK} and Gilbert theorem \cite{Gilbert} for $\wb$-ensembles we assume that the first $r$ eigenstates of $\hat{H}(\hat{h})$ are non-degenerate. In that case we have for \emph{every} $\wb$ with at most $r$ non-vanishing weights the following sequence of three consecutive maps,
\begin{equation}\label{seq-wRDMFT}
\mbox{$\wb$-RDMFT:}\quad \hat{h} \xmapsto{(1)} \hat{H}(\hat{h}) \xmapsto{(2)} \Gwb(\hat{h}) \xmapsto{(3)} \gwb(\hat{h})\,,
\end{equation}
where $\gwb(\hat{h}) \equiv N \mbox{Tr}_{N-1}[\Gwb(\hat{h})]$ denotes the 1RDM of the $\wb$-minimizer $\Gwb(\hat{h})$ of the Hamiltonian $\hat{H}(\hat{h})$. We call a 1RDM $\gwb$ $\wb$-ensemble $v$-representable \footnote{Although the variable part of the Hamiltonian is given by the full one-particle Hamiltonian $\hat{h}$ rather than the external potential $\hat{v}$ we still use here the established term ``$v$-representability'', as in the context of ground state RDMFT and DFT.} if there exists a corresponding $\hat{h}$ such that $\gwb$ follows from  the corresponding sequence \eqref{seq-wRDMFT}, i.e., $\gwb=\gwb(\hat{h})$.

To discuss the possible validity of a generalized Hohenberg-Kohn theorem in $\wb$-ensemble RDMFT, we observe that the map (1), $\hat{h} \mapsto \hat{h} +\hat{W}$, is apparently invertible. In the context of DFT,
\begin{equation}\label{seq-DFT}
\mbox{DFT:}\quad \hat{v} \xmapsto{(1)} \hat{H}(\hat{v}) \xmapsto{(2)} \Gh(\hat{v}) \xmapsto{(3)} \rho(\hat{v}) \,,
\end{equation}
it is exactly the Hohenberg-Kohn theorem which proves also the invertibility of the maps (2) and (3) \cite{HK}, and thus we have in particular $\Gh \equiv \Gh(\rho)$. Since the $N$-particle quantum state $\Gh$ determines all properties of its quantum system, already the particle density $\rho$ alone determines, at least in principle, various physical properties of the system with Hamiltonian $\hat{H}(\hat{v})$ in its ground state. The analogous statement can also be proven in the context of ground state RDMFT, i.e., for \eqref{seq-wRDMFT} with $\wb=\wb_0\equiv (1,0,\ldots,0)$, with the 1RDM $\gh$ rather than the density $\rho$ as natural variable \cite{Gilbert}. Yet, map (2) cannot be inverted anymore since it is a many-to-one map. For instance, this can be deduced from finite fermionic or bosonic lattice systems with $\hat{t}=0$: The respective eigenstates are given by the configuration states, where each particle occupies a lattice site. Accordingly,  the ordering of the energy eigenvalues is insensitive to a sufficiently small change of the external potential (see also, e.g., Ref.~\cite{Benavides20}) and thus each configuration state appears as ground state for multiple one-particle Hamiltonians.

Since ground state RDMFT is contained in our $\wb$-ensemble RDMFT, or by generalizing the previous argument to $\wb$-ensembles, we conclude that map (2) will not be invertible in $\wb$-ensemble RDMFT. In contrast, map (3) is still invertibility which we prove in the following by \emph{reductio ad absurdum}. For this, we assume that there existed two different $\wb$-ensemble minimizers $\Gwb^{(1,2)}$ with the same 1RDM $\gwb$. Since this assumption refers to the framework \eqref{seq-wRDMFT} of $\wb$-ensemble RDMFT, there would then exist two different $\hat{h}^{(1,2)}$ such that $\Gwb^{(1,2)}$ would be the unique minimizer \eqref{Gamma_minimizer} for $\hat{H}(\hat{h}^{(1,2)})$ in \eqref{GOK}. Hence, we would have
\begin{widetext}
\begin{equation}\label{HKproof1}
E_{\wb}(\hat{h}^{(1)}) = \mbox{Tr}_N\big[\hat{H}(\hat{h}^{(1)})\Gwb^{(1)}\big] < \mbox{Tr}_N\big[\hat{H}(\hat{h}^{(1)})\Gwb^{(2)}\big]= \mbox{Tr}_N\big[\hat{H}(\hat{h}^{(2)})\Gwb^{(2)}\big] + \mbox{Tr}_1\big[\big(\hat{h}^{(1)}-\hat{h}^{(2)}\big)\gwb\big]\,.
\end{equation}
\end{widetext}
Repeating the same derivation for interchanged indices, $1 \leftrightarrow 2$, and then adding the two inequalities \eqref{HKproof1} would yield the contradiction $E_{\wb}(\hat{h}^{(1)})+ E_{\wb}(\hat{h}^{(2)}) < E_{\wb}(\hat{h}^{(1)})+ E_{\wb}(\hat{h}^{(2)})$. Accordingly, in the context \eqref{H} of Hamiltonians of interest and for any $\wb$ there exists a corresponding one-to-one correspondence between $\wb$-ensemble minimizer states $\Gwb$ and $\wb$-ensemble $v$-representable 1RDMs $\gwb$. This also implies that the expectation value of any observable of a system in a $\wb$-ensemble minimized state can be understood as a functional of the underlying 1RDM. In particular, this applies to the $\wb$-ensemble energy $E_{\wb}$ and its one- and two-particle interaction contributions. 

In order to prove the analogue of the second part of the Hohenberg-Kohn theorem, we observe, as a direct consequence of the variational principle \eqref{GOK},
\begin{equation}
E_{\wb}(\hat{h}) = \mbox{Tr}_N[\hat{H}(\hat{h})\Gwb(\hat{h})] \leq \mbox{Tr}_N[\hat{H}(\hat{h})\Gwb(\hat{h}')]
\end{equation}
for any possible $\wb$-ensemble minimizer $\Gwb(\hat{h}')$ \eqref{Gamma_minimizer}.
Exploiting then for $\wb$-ensemble $v$-representable 1RDMs $\gwb$ their one-to-one correspondence to $\wb$-ensemble minimizers $\Gwb$ yields (with $\gwb \equiv \gwb(\hat{h})$)
\begin{equation}\label{HK2}
E_{\wb}(\hat{h}) = \mbox{Tr}_1[\hat{h}\gwb] + \Fw(\gwb) \leq \mbox{Tr}_1[\hat{h}\gwb'] + \Fw(\gwb')\,,
\end{equation}
where $\Fw(\gwb) \equiv \mbox{Tr}_N[\hat{W}\Gwb(\gwb)]$ is the universal functional, defined on the domain of $\wb$-ensemble $v$-representable 1RDMs. Equation \eqref{HK2} means nothing else than that the $\wb$-weighted energy $E_{\wb}(\hat{h})$ of $\hat{H}(\hat{h})$ can be obtained variationally, namely by minimizing $\mbox{Tr}_1[\hat{h}\gwb'] + \Fw(\gwb')$. Moreover, the minimum is attained for the true physical 1RDM $\gwb(\hat{h})$.

We conclude this section, by presenting an equation for calculating indirectly the universal functional. It refers to the sequence \eqref{seq-wRDMFT}: For any $\wb$-ensemble $v$-representable 1RDM $\gwb \equiv \gwb(\hat{h})$, the generalized Hohenberg-Kohn theorem implies
\begin{equation}
\Fw(\gwb) = E_{\wb}(\hat{h}) - \mbox{Tr}_1[\hat{h}\gwb]\,.
\end{equation}
Due to the prospects of machine learning techniques and big data science, this relation is gaining relevance. For instance, it has been used recently to determine in ground state DFT of one-dimensional systems a rather accurate approximation to the exchange-correlation functional. For this, the ground state problem was solved for numerous external potentials with the help of the density matrix renormalization group ansatz and this data has then be exploited in some fitting scheme \cite{Lubasch2016}.

\section{Constrained search formalism for $\bd w$-ensembles \label{sec:wRDMFT-LL}}
The results of Sec.~\ref{sec:wRDMFT} establish $\wb$-ensemble RDMFT on a rather abstract level. It is not clear yet how this approach based on the generalization of the Hohenberg-Kohn theorem could be realized in practice. In particular, one would first need to solve the corresponding  intricate $\wb$-ensemble $v$-representability problem to determine the functional's domain. Moreover, restricting to spin-symmetry sectors does not necessarily remove all energy degeneracies of realistic systems and thus our generalized Hohenberg-Kohn theorem would not always be applicable. To deal with these caveats, we work out in the following the same ideas that were quite effective for the development of DFT and RDMFT for ground states.

\subsection{Notation and constrained search for ground states}\label{sec:RDMFT-LL}
We first introduce some notation and briefly recap some key results on ground state RDMFT. These concepts will reappear in the following sections in a modified form, adapted to the description of excited states. Since we are interested in describing $N$-particle quantum systems, we first introduce the $N$-particle Hilbert space $\mathcal{H}_N$ and denote by $\mathcal{H}_1$ the underlying $d$-dimensional one-particle Hilbert space. For identical fermions, all states in $\mathcal{H}_N$ are antisymmetric under the exchange of two particles, $\mathcal{H}_N \equiv \wedge^N[\mathcal{H}_1]$, whereas for bosons they are symmetric, $\mathcal{H}_N \equiv \mathcal{S}^N[\mathcal{H}_1]$. Moreover, we denote by $\mathcal{E}^N$ the set of all ensemble $N$-particle density operators $\hat{\Gamma}$. Then, the extremal elements of $\mathcal{E}^N$ form the set $\mathcal{P}^N$ of pure states $\hat{\Gamma}\equiv\ket{\Psi}\!\bra{\Psi}$. By tracing out $N-1$ particles from a state $\hat{\Gamma}\in \mathcal{E}^N$, we obtain the one-particle reduced density operator $\hat{\gamma}$,
\begin{equation}\label{1RDM}
\hat{\gamma}\equiv N\Tr_{N-1}[\hat{\Gamma}] = \sum_{i=1}^d\lambda_i\ket{i}\!\bra{i}\,,
\end{equation}
where the eigenvalues $\lambda_i$ are the so-called natural occupation numbers and $\ket{i}$ the corresponding natural orbitals.

According to Levy \cite{LE79} and Lieb \cite{L83}, the ground state energy $E(\hat{h})$ and the ground state 1RDM then follow from the Rayleigh-Ritz variational principle as
\begin{eqnarray}\label{Levy}
E(\hat{h}) &=& \min_{\hat{\Gamma}\in \mathcal{P}^N} \Tr_N\left[(\hat{h}+\hat{W})\hat{\Gamma}\right]\nonumber\\
&\equiv& \min_{\hat{\gamma}\in \mathcal{P}^1_N}\left[\Tr_1[\hat{h}\hat{\gamma}] + \mathcal{F}(\hat{\gamma})\right]\,.
\end{eqnarray}
The pure ground state universal functional $\mathcal{F}(\hat{\gamma})$ in this so-called constrained search formalism follows from the minimization of the expectation value $\Tr_N[\hat{W}\hat{\Gamma}]$ over all $\hat{\Gamma}\in \mathcal{P}^N$,
\begin{equation}
\mathcal{F}(\hat{\gamma}) \equiv \min_{\mathcal{P}^N\ni\hat{\Gamma}\mapsto\hat{\gamma}}\Tr_N[\hat{W}\hat{\Gamma}]\,.
\end{equation}
In its original formulation by Levy \cite{LE79}, the minimization in Eq.~\eqref{Levy} is performed over all pure states $\hat{\Gamma}\in \mathcal{P}^N$ and thus the domain of $\mathcal{F}$ consists of all pure state $N$-representable 1RDMs $\hat{\gamma}\in\mathcal{P}^1_N= N\Tr_{N-1}[\mathcal{P}^N]$. However, the non-convex set $\mathcal{P}^1_N$ is in general unknown for fermions, a fact which is due to the too complicated generalized Pauli constraints \cite{KL06,AK08,S18atoms}.
Justified by the observation that an exact convex relaxation \cite{rockafellar2015} of the non-convex optimization problem \eqref{Levy} does not change the outcome of the minimization, Valone \cite{V80} proposed to extend the domain of $\mathcal{F}$ to all ensemble $N$-representable $\hat{\gamma}\in \mathcal{E}^1_N= N\Tr_{N-1}[\mathcal{E}^N]$.
Indeed, both sets $\mathcal{E}^N$ and $\mathcal{E}^1_N$ are convex and in particular $\mathcal{E}^N_1 = \mathrm{conv}(\mathcal{P}^N_1)$, where $\mathrm{conv}(\cdot)$ denotes the convex hull (see Ref.~\cite{rockafellar2015}). Moreover, the corresponding ensemble ground state functional $\xbar{\F}$ is equal to the lower convex envelope of $\F$ \cite{Schilling2018},
\begin{equation}
\xbar{\F}(\hat{\gamma}) = \mathrm{conv}\left(\F(\hat{\gamma})\right)\,.
\end{equation}

For fermions, Valone's \cite{V80} formulation of RDMFT is the crucial step which turned RDMFT into a practical method since the set $\mathcal{E}^1_N$ is only restricted through the well-known Pauli exclusion principle $0\leq \lambda_i\leq 1$. In the case of bosons, the natural occupation numbers are only restricted through $0\leq \lambda_i\leq N$ and there are no additional constraints as the generalized Pauli constraints for fermions. Therefore, the pure state $N$-representability constraints for bosons are known and, in particular, $\mathcal{E}^1_N=\mathcal{P}^1_N$ \cite{GR19,Benavides20}. Nevertheless,
relaxing the minimization in Eq.~\eqref{Levy} to a convex one is also advantageous for bosons since any local minimum of a convex functional is also a global minimum.


\subsection{Extension of constrained search to $\bd w$-ensembles\label{sec:foundations}}

The GOK variational principle \eqref{GOK} applied to the $N$-particle Hamiltonian $\hat{H}(\hat{h})$ \eqref{H} in combination with the constrained search leads to an $\wb$-ensemble functional in a straightforward manner:
\begin{equation}\label{Levy_w}
\begin{split}
E_{\boldsymbol{w}}(\hat{h}) &= \min_{\hat{\Gamma}\in \mathcal{E}^N(\boldsymbol{w})}\Tr_N[(\hat{h} + \hat{W})\hat{\Gamma}]\\\
&= \min_{\hat{\gamma}\in \mathcal{E}_N^1(\boldsymbol{w})}\Big[\min_{\mathcal{E}^N(\boldsymbol{w})\ni\hat{\Gamma}\mapsto\hat{\gamma}}\Tr_N[(\hat{h}+\hat{W})\hat{\Gamma}]\Big] \\\
&= \min_{\hat{\gamma}\in \mathcal{E}_N^1(\boldsymbol{w})}\Big[\Tr_1[\hat{h}\hat{\gamma}] + \mathcal{F}_{\boldsymbol{w}}(\hat{\gamma})\Big]\,,
\end{split}
\end{equation}
where
\begin{equation}\label{Fw_def}
\mathcal{F}_{\boldsymbol{w}}(\hat{\gamma})\equiv \min_{\mathcal{E}^N(\boldsymbol{w})\ni\hat{\Gamma}\mapsto\hat{\gamma}}\Tr_N[\hat{W}\hat{\Gamma}]\,.
\end{equation}
It is worth noticing that for $\boldsymbol{w}_0 \equiv (1, 0, \ldots)$ we recover ground state RDMFT and \eqref{Levy_w} reduces indeed to \eqref{Levy}.
Moreover, Eq.~\eqref{Levy_w} defines the universal $\bd w$-ensemble functional $\Fw(\hat{\gamma})$ in a similar fashion as Eq.~\eqref{Levy} defines the universal ground state functional $\mathcal{F}$.
The domain of $\Fw$ is given by those 1RDMs $\hat{\gamma}\in\ew$ which follow from an $N$-boson density operator $\hat{\Gamma}\in\Ew$ by tracing out $N-1$ particles. Thus, an $\boldsymbol{w}$-ensemble $N$-representability problem arises.
Due to the nonlinear spectral restriction of $\mathcal{E}^N$ to $\mathcal{E}^N(\boldsymbol{w})$, both sets $\Ew$ and $\ew$ are not convex in striking contrast to $\mathcal{E}^N$ and $\mathcal{E}^1_N$, respectively. In analogy to Levy's ground state RDMFT,
the description of $\mathcal{E}^1_N(\bd w)$ would involve for any $\bd w$ highly involved additional constraints on the natural occupation numbers.

\subsection{Convex relaxation of $\bd{w}$-ensemble RDMFT\label{sec:relaxation}}

Since the $\bd w$-ensemble $N$-representability problem is too intricate, the excited state RDMFT introduced in Sec.~\ref{sec:foundations} is not feasible from a practical point of view: Without knowing the functional's domain $\ew$, the process of deriving functional approximations cannot be initiated. Yet, any non-convex minimization problem can be turned into a corresponding convex one, at least in principle \cite{rockafellar2015}. This crucial observation from convex analysis allows us to circumvent the too involved $\wb$-ensemble $N$-representability constraints in the same way as Valone's approach circumvented the generalized Pauli constraints (recall Sec.~\ref{sec:foundations}). Applied to the constrained search formalism in Eq.~\eqref{Levy_w}, this means to replace the universal  $\bd w$-ensemble functional $\Fw(\hat{\gamma})$ by its lower convex envelope,
\begin{equation}\label{Fbw}
\Fbw(\hat{\gamma})\equiv\mathrm{conv}\left(\mathcal{F}_{\boldsymbol{w}}(\hat{\gamma})\right)\,,
\end{equation}
whose domain follows as the convex hull of $\mathcal{E}^1_N(\bd w)$,
\begin{equation}
\ebw \equiv \mathrm{conv} \left(\mathcal{E}_N^1(\boldsymbol{w})\right)\,.
\end{equation}
Since the partial trace map $\Tr_{N-1}[\cdot]$ is linear, it commutes with the convex hull operation $\mathrm{conv}(\cdot)$. This allows us to obtain a more concrete characterization of the convex hull of $\ew$  (the proof is identical to the one of the same statement for fermions as presented in \cite{LCLS21})
\begin{equation}\label{eq:EN1union}
\ebw = N\Tr_{N-1}[\Ebw]=\bigcup\limits_{\bd w^\prime\prec\,\bd w}\mathcal{E}^1_N(\bd w^\prime)\,.
\end{equation}
Here, a vector $\bd w^\prime\in \RR^D$ is majorized by a vector $\bd w\in \RR^D$, $\bd w^\prime\prec\,\bd w$, if and only if
\begin{equation}\label{major}
w_1^{\prime \downarrow} + \ldots+ w_k^{\prime \downarrow}\leq w_1^\downarrow + \ldots + w_k^\downarrow \quad \forall \,1\leq k\leq D\,,
\end{equation}
with equality for $k=D$. The arrow $\downarrow$ indicates that \eqref{major} refers to the vector entries arranged in decreasing order.
In the following, we can always omit this superscript since the entries of weight vectors $\bd w$ are already ordered decreasingly by definition.
Moreover, the last equality in Eq.~\eqref{eq:EN1union} leads to the following inclusion relation,
\begin{equation}\label{inclusion}
\bd w^\prime\prec\bd w\,\,\Leftrightarrow\,\,\xbar{\mathcal{E}}^1_N(\bd{w}^\prime)\subset\ebw\,.
\end{equation}
Hence, the smaller the weight vector $\wb$ with respect to majorization, the smaller will be the set of relaxed $\bd w$-ensemble $N$-representable 1RDMs, i.e.~the domain of $\Fbw$. We will illustrate this inclusion relation for several weight vectors $\bd w$ in Fig.~\ref{fig:inclusion}, where the shrinking volume of the respective domains is in evidence.

Clearly, the convex relaxation does not change the outcome for the energy $E_{\bd w}$ and, in particular,
\begin{equation}
E_{\bd w}=\min_{\hat{\gamma}\in \ebwS}\left[\Tr_1[\hat{h}\hat{\gamma}] + \Fbw(\hat{\gamma})\right]\,.
\end{equation}
It is worth noticing that the same relaxed $\bd w$-ensemble RDMFT can be obtained by starting at the $N$-boson level, where $\Ebw = \mathrm{conv}(\Ew)$. Thus, replacing the set $\Ew$ by its convex hull in the constrained search formalism \eqref{Levy_w} immediately leads to a more concrete expression for the relaxed $\bd w$-ensemble functional 
\begin{equation}\label{Fbw_min}
\Fbw(\hat{\gamma}) = \min_{\EbwS \ni\hat{\Gamma}\mapsto \hat{\gamma}} \Tr_N[\hat{\Gamma} \hat{W}]\,.
\end{equation}
The importance of the constrained search expression \eqref{Fbw_min} of the relaxed $\bd w$-ensemble functional $\Fbw$ can hardly be overestimated. It will namely serve as the starting point for the construction of functional approximations.

\subsection{Determining the functional's domain\label{sec:characterization}}
A formal definition of the set $\ebw$ as in Sec.~\ref{sec:relaxation} without a constructive approach to characterize it is apparently not sufficient for practical purposes. It would be unclear how to derive approximations of $\Fbw$ and how to minimize then the total energy functional $\Tr_1[h\gamma]+\Fbw$ \eqref{Fbw_min} over the set $\ebw$. Hence, Eq.~\eqref{Fbw_min} only leads to a viable $\bd w$-ensemble RDMFT if a compact description of the functional's domain is found.
In the following, we derive the vertex representation of the functional's domain which takes effectively the form of a polytope.  To this end, we will exploit a fruitful analogy between the mathematical problem of determining the set $\ebw$ and the description of non-interacting bosons. In particular, this will allow us to resort to our physical intuition about non-interacting bosons, leading to a better understanding of the boundary of $\ebw$.
\begin{figure}[htb]
\includegraphics[width=\linewidth]{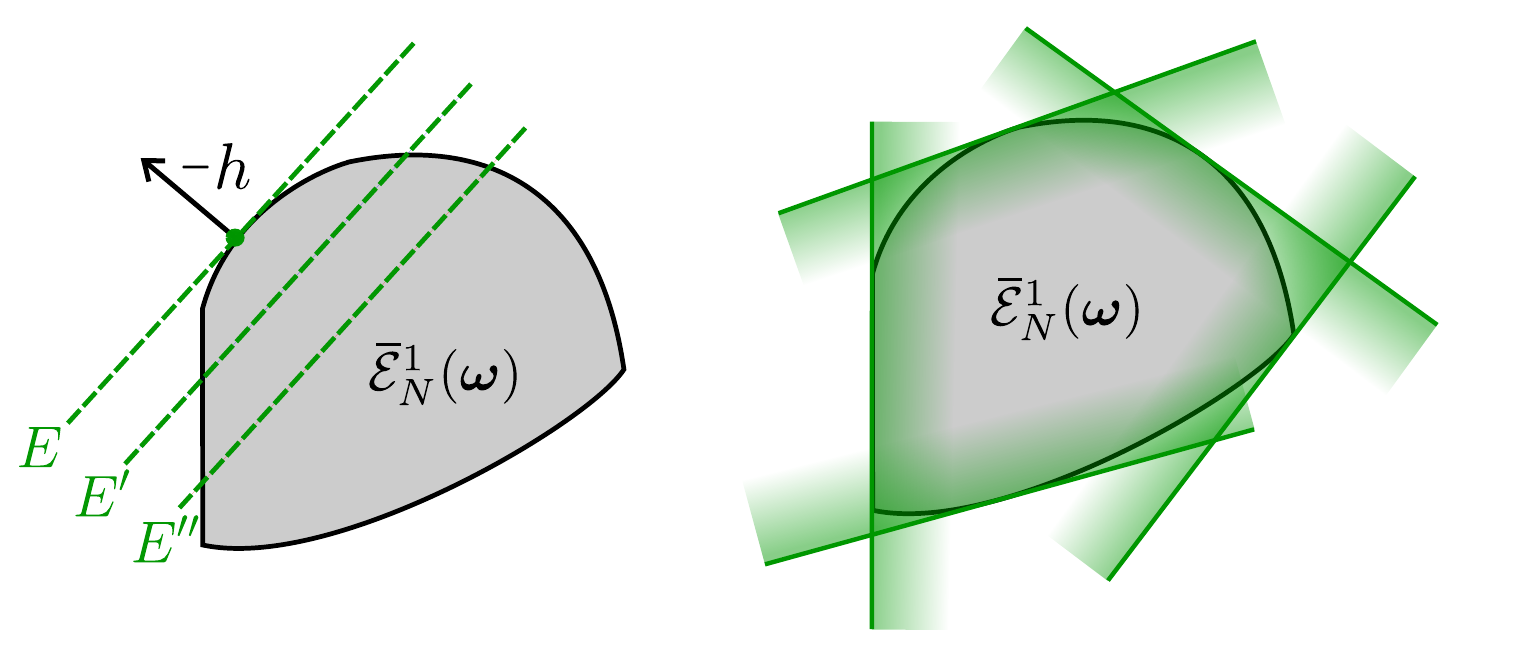}
\caption{Illustration of the duality correspondence. Left: Schematic illustration of the minimization of $\mbox{Tr}_1[h (\cdot)]$ over a compact convex set $\ebw$. Right: Performing the minimization for all possible ``directions" $h$ determines the boundary of $\ebw$. \label{fig:dual}}
\end{figure}

\begin{figure*}[htb]
\includegraphics[width=.35\linewidth]{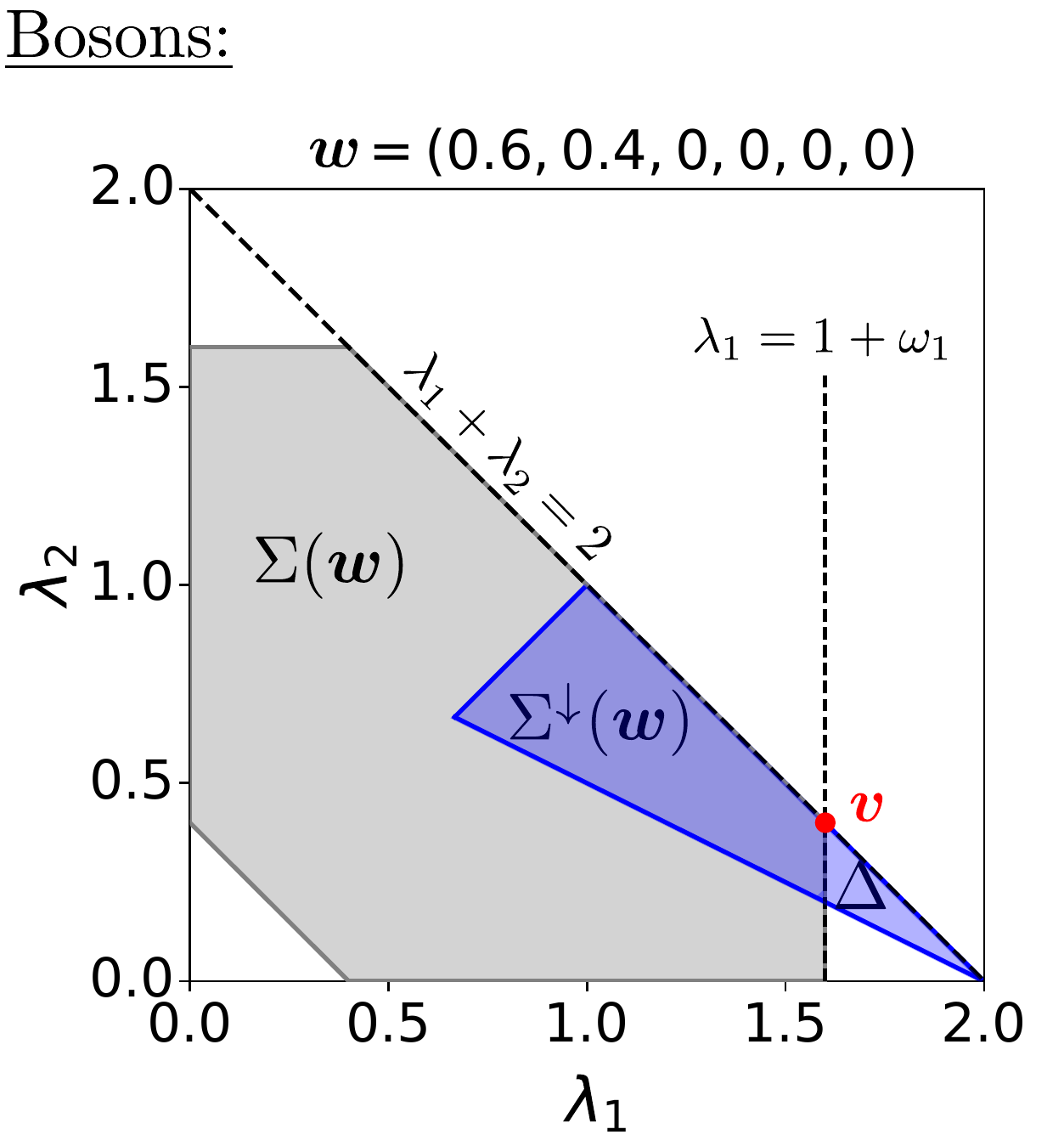}
\hspace{3cm}
\includegraphics[width=.35\linewidth]{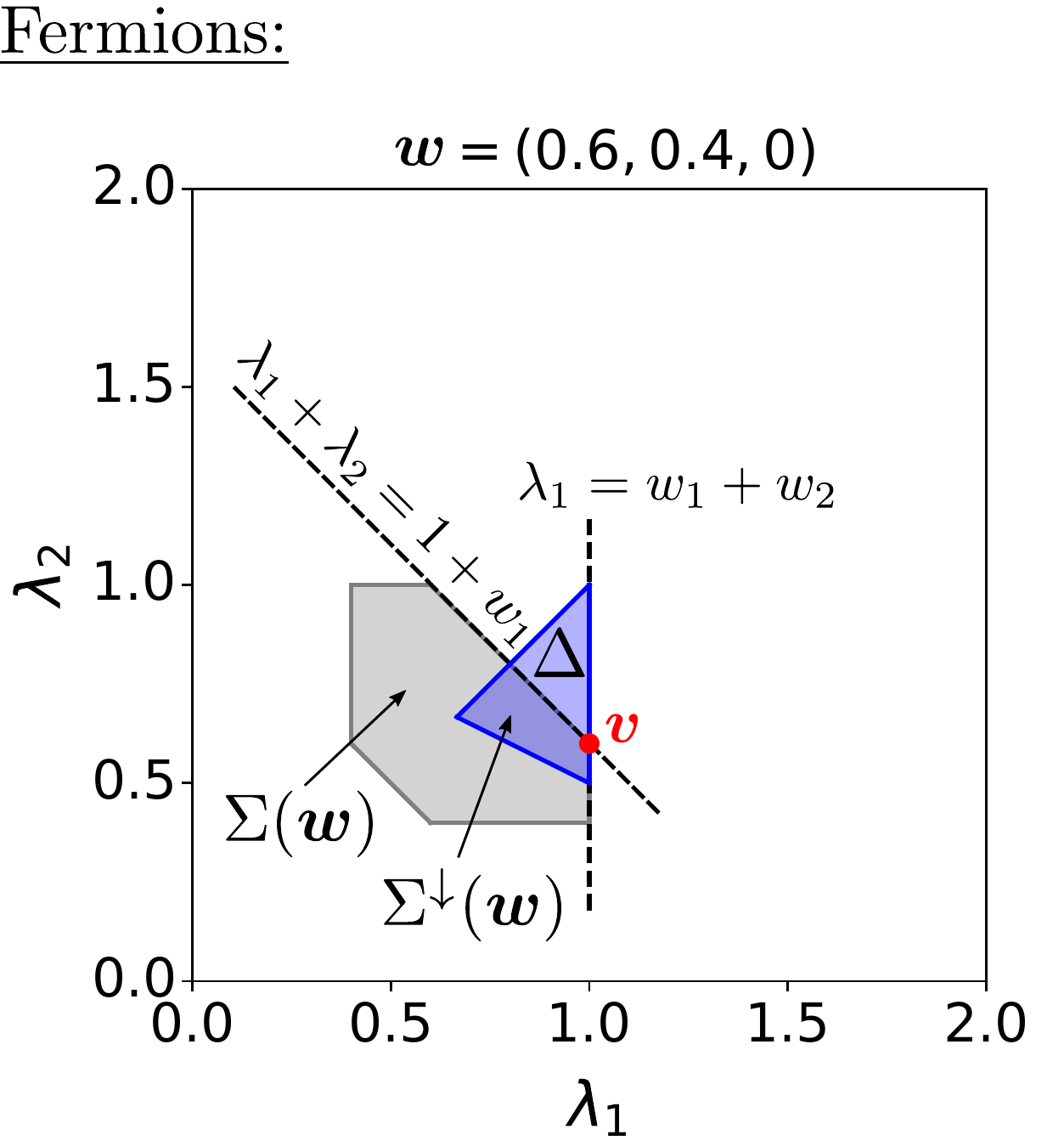}
\caption{Left: Illustration of the bosonic spectral polytopes $\Sigma(\boldsymbol{w})$ (grey) and $\Sigma^\downarrow(\boldsymbol{w})$ (overlap between grey and blue) and $\Delta$ (blue) for $(N, d)=(2, 3)$. (See text for more explanations.) Right: Illustration of the fermionic spectral polytopes for the setting $(N, d)=(2, 3)$ introduced in Ref.~\cite{Schilling21, LCLS21}. Note that the different length of the weight vector $\bd w$ is due to the different dimension of the two particle Hilbert space for bosons and fermions. \label{fig:sigma} }
\end{figure*}

According to a duality principle \cite{rockafellar2015}, every compact convex set can be characterized equivalently either through all its points or all its supporting hyperplanes. In the following, as it is illustrated in Fig.~\ref{fig:dual}, we apply this to our compact convex set $\ebw$. First, since $\Tr_1[\hat{h}\hat{\gamma}]\equiv \langle \hat{h}, \hat{\gamma}\rangle_1$ defines an inner product, a notion of geometry exists on the underlying space of 1RDMs. Accordingly, any one-particle Hamiltonian $\hat{h}$ defines a direction within this space.
Minimizing $\langle \hat{h}, \hat{\gamma}\rangle_1$ over all 1RDMs $\hat{\gamma}\in \ebw$ corresponds to shifting the hyperplane of constant value $\langle \hat{h}, \hat{\gamma}\rangle_1$ in direction $-\hat{h}$ until it touches the boundary of $\ebw$. Performing this minimization for all one-particle Hamiltonians $\hat{h}$ determines $\ebw$ entirely.
Since the spectral vector $\boldsymbol{w}$ refers primarily to the $N$-boson level, we lift the minimization to the $N$-boson level through
\begin{equation}\label{lifting}
\min_{\hat{\gamma}\in\ebwS}\Tr_1[\hat{h}\hat{\gamma}]= \min_{\hat{\Gamma}\in \EbwS}\Tr_N[\hat{h}\hat{\Gamma}]\,.
\end{equation}
Furthermore, we observe that the set $\ebw$ is invariant under unitary transformations,
\begin{equation}\label{unitaryinv}
\hat{u}\ebw \hat{u}^\dagger=\ebw\,,
\end{equation}
for all unitaries $\hat{u}$ acting on the one-particle Hilbert space $\mathcal{H}_1$. Accordingly, we can restrict our procedure of characterizing $\ebw$ through supporting hyperplanes to one-particle Hamiltonians $\hat{h}$ with an arbitrary but fixed eigenbasis, $\hat{h}=\sum_{i=1}^dh_i\ket{i}\!\bra{i}$, and $h_1\leq h_2\leq \ldots\leq h_d$.
Equivalently, the knowledge of the spectrum of a 1RDM $\hat{\gamma}$ is sufficient to determine whether $\hat{\gamma}$ belongs to the set $\ebw$ or not. Accordingly, we are interested in describing the set of all admissible spectra $\bd{\lambda} \equiv \mathrm{spec}(\hat{\gamma})$,
\begin{equation}
\Sigma(\bd w)\equiv \mathrm{spec}\left(\ebw\right)\,.
\end{equation}
As our following derivation reveals, $\Sigma(\bd w)$ takes the form of a polytope, i.e., it is the convex hull of a finite number of vertices $\bd{v}^{(l)}$. Furthermore, in agreement with \eqref{unitaryinv}, the spectral polytope $\Sigma(\bd w)\subset \RR^d$ is invariant under permutations of the Cartesian coordinates. Without loss of generality we therefore focus for a moment on the set $\Sigma^\downarrow(\bd{w})$ of decreasingly ordered vectors
\begin{equation}\label{SdD}
\Sigma^\downarrow(\bd{w}) = \Sigma(\bd w)\cap \Delta\,,
\end{equation}
where $\Delta$ denotes the set
\begin{equation}
\Delta = \{\boldsymbol{\lambda}^\downarrow\in \RR^d\,|\,N\geq\lambda_1^\downarrow\geq \lambda_2^\downarrow\geq \ldots\geq \lambda_d^\downarrow\geq 0\}\,.
\end{equation}
In contrast to the fermionic case \cite{Schilling21}, the natural occupation numbers $\lambda_i^\downarrow$ in the set $\Delta$ do not obey the Pauli exclusion principle. This in turn results in a different underlying combinatorial structure leading to a qualitative different characterization of $\Sigma^\downarrow(\bd{w})$ than in the fermionic case \cite{LCLS21}.
Before presenting the general procedure for calculating $\Sigma^\downarrow(\bd{w})$, we illustrate in the left panel of Fig.~\ref{fig:sigma} relation \eqref{SdD} between the three sets $\Sigma(\bd w)$ (grey), $\Sigma^\downarrow(\bd w)$ (overlap between grey and blue) and $\Delta$ (blue) for $(N, d)=(2, 3)$. Due to the normalization of $\bd\lambda$ to the total particle number, we can omit the third natural occupation number $\lambda_3$, and focus solely on $\lambda_1$ and $\lambda_2$. The spectral polytope $\Sigma^\downarrow(\bd w)$ is restricted through the two equalities on the natural occupation numbers $\lambda_1$ and $\lambda_2$ indicated by the dashed lines (see Sec.~\ref{sec:r2} for their derivation). The red point marks the position of the single independent vertex $\bd v$ (the other five vertices follow from $\bd{v}$ by permutation of its entries). We provide a systematic derivation of $\Sigma^\downarrow(\bd w)$ and the vertex $\bd v$ in the following. To illustrate the differences to the fermionic case discussed in Ref.~\cite{Schilling21, LCLS21}, we show in the right panel of Fig.~\ref{fig:sigma} the spectral polytopes for $N=2$ fermions in $d=3$ orbitals and the same choice of the weight vector $\bd w$.

To calculate the vertices $\bd v^{(l)}$ of the bosonic spectral polytope $\Sigma^\downarrow(\bd w)$, we start by determining for every set of energy levels $\bd h\equiv \bd h^\uparrow \equiv (h_1, \ldots, h_d)$ the sequence
\begin{equation}\label{sequence}
\bd h\mapsto \hat{\Gamma} \mapsto \hat{\gamma}\mapsto \bd v\,.
\end{equation}
Here, $\hat{\Gamma}$ denotes the minimizer state of the right side of Eq.~\eqref{lifting} and $\bd v = \mathrm{spec}(\hat{\gamma})$ the spectrum of the resulting 1RDM $\hat{\gamma}$. The eigenstates of a one-particle Hamiltonian $\hat{h}$, represented by the vector $\bd h$, on the $N$-boson Hilbert space are given by the configuration states $\ket{\bd i} \equiv \ket{i_1, \ldots, i_N}$.
To determine in a systematic manner all distinct vertices $\bd v$ in Eq.~\eqref{sequence}, we first introduce a partial ordering of configurations $\bd i$, where a configuration is an element of the set
\begin{equation}\label{conf_set}
\mathcal{I}_{N, d} \equiv \{ \bd{i}\equiv(i_1, \ldots, i_N)|\,1\leq i_1\leq i_2\leq \ldots\leq i_N\leq d\}\,.
\end{equation}
Furthermore, we can assign an energy to every configuration through $\bd h$ and thus order different configurations according to their energy.
Then, for any two configurations $\bd i$ and $\bd j$ we define the partial ordering
\begin{eqnarray}\label{partial}
\bd i\leq \bd j\quad:\Leftrightarrow &\quad &  \sum_{k=1}^Nh_{i_k}\leq  \sum_{k=1}^Nh_{j_k}\quad \forall \,\bd h\equiv \bd h^\uparrow \nonumber \\
\Leftrightarrow &\quad & i_k\leq j_k\quad \forall\,1\leq k\leq N\,.
\end{eqnarray}
Hence, the partial order of configurations in $\mathcal{I}_{N, d}$ is completely characterized by the eigenenergies of a chosen one-particle Hamiltonian $\hat{h}$.
The partial ordering \eqref{partial} leads for each choice of $\bd h$ to a corresponding lineup $l$,
\begin{equation}\label{lineup}
\bd i_1\to \bd i_2 \to \ldots\to \bd i_r\,,
\end{equation}
of the energetically lowest configurations $\bd i_j$. Due to reasons that become clear in the following, the length of those lineups is restricted to the number $r$ of non-vanishing weights $w_j$ in the weight vector $\bd w$.
\begin{figure*}
\includegraphics[width=0.4\linewidth]{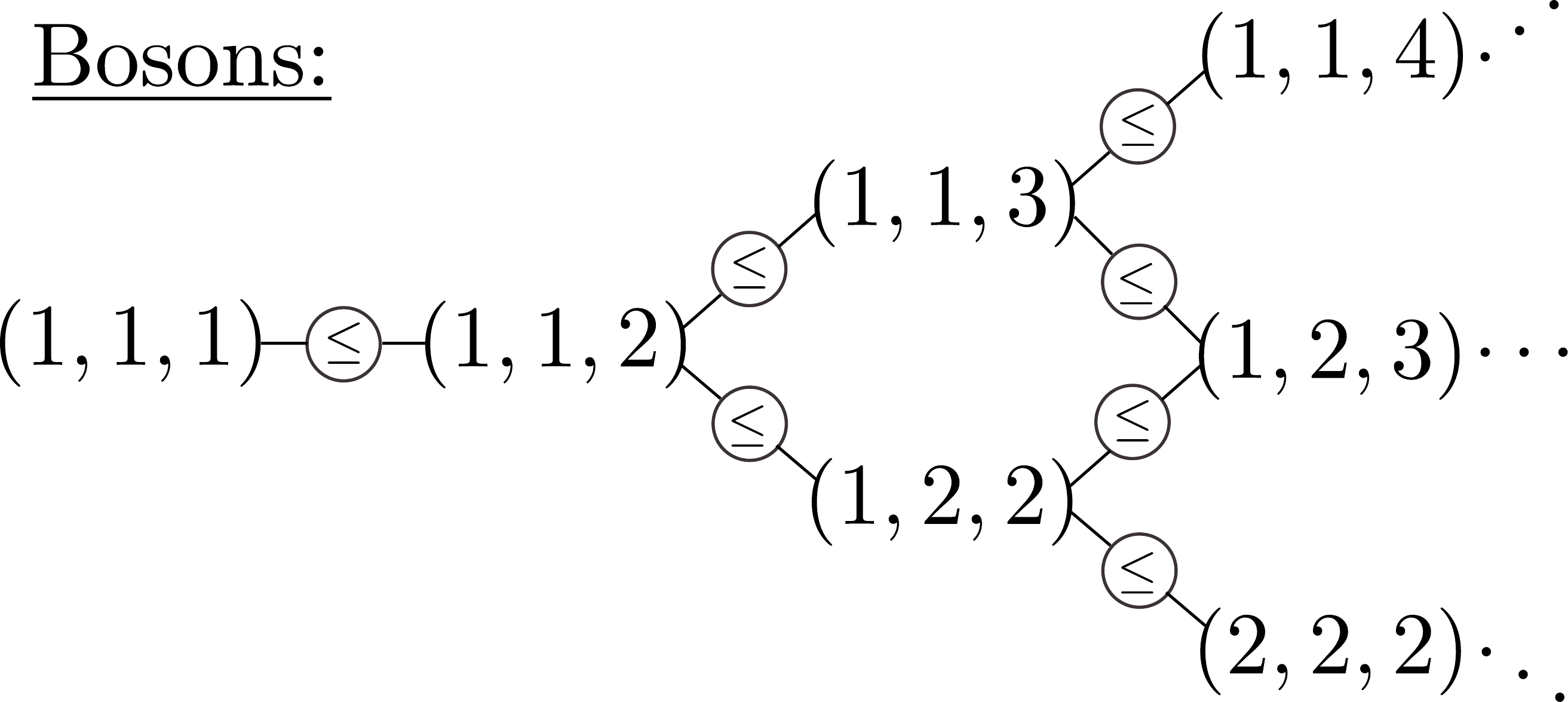}
\hspace{2cm}
\includegraphics[width=0.4\linewidth]{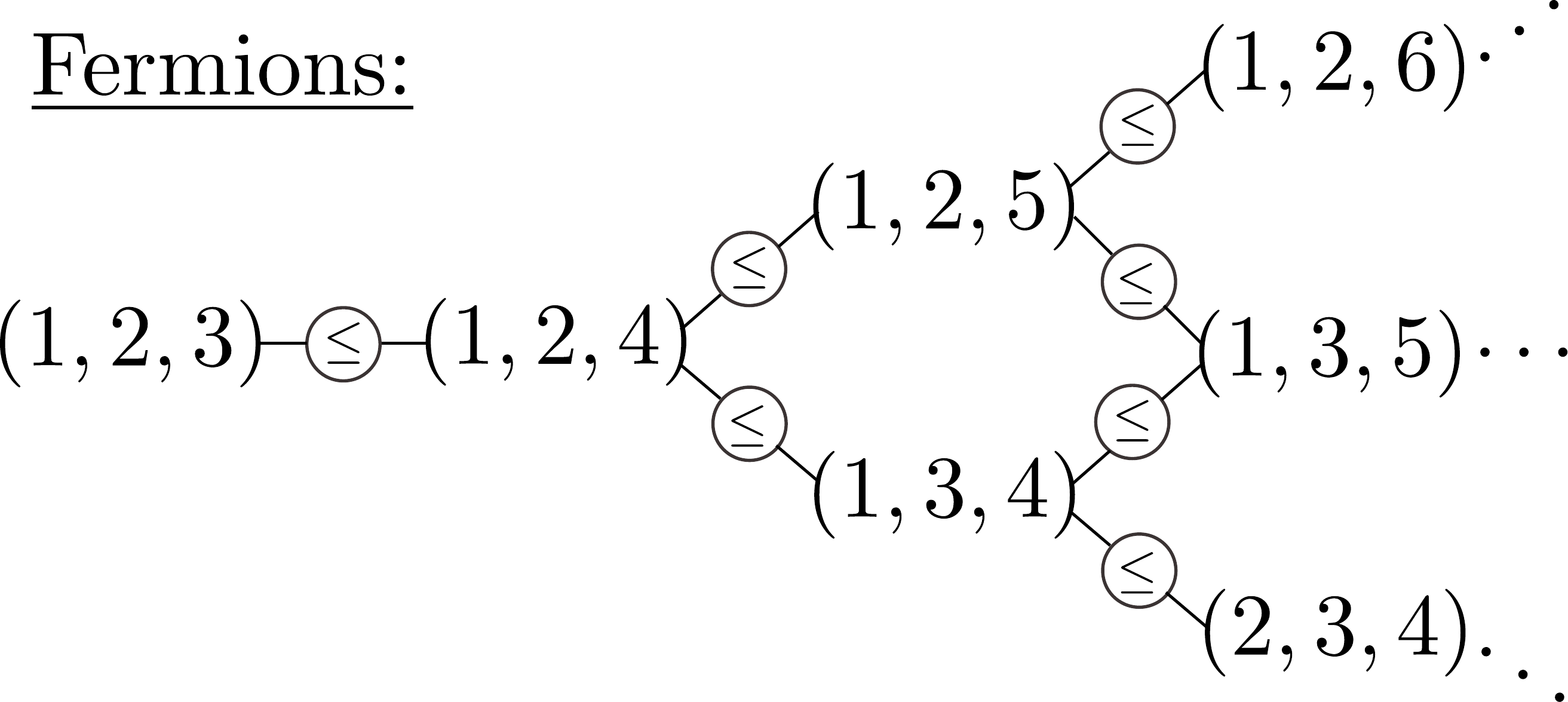}
\caption{Illustration of the structure of the ``excitation spectrum'' for $N=3$ bosons (left) and $N=3$ fermions (right). See text for more details. \label{fig:spectrum}}
\end{figure*}
As an illustration, we show in the left panel of Fig.~\ref{fig:spectrum} the structure of this ``excitation spectrum'' for $N=3$ bosons following from the partial ordering of configurations in Eq.~\eqref{partial}. For $r=1$, the lineup \eqref{lineup} consists of only one configuration $(1,1,1)$. For $r=2$, we have again a single lineup $(1,1,1)\to(1,1,2)$ now consisting of two configurations. Note that according to \eqref{GOK}, $\wb$-RDMFT with $r=2$ is already sufficient for calculating the ground state energy and its gap to the first excited state. For $r=3$ there are now two choices of the second excitation leading to two lineups.
Furthermore, for larger values $r\leq 4$, both configurations $(1,1,3)$ and $(1,2,2)$ must appear before $(1,2,3)$, a pattern which can be easily continued to derive all distinct lineups \eqref{lineup} required to calculate the vertex representation of $\Sigma(\bd w)$.
The right panel of Fig.~\ref{fig:spectrum} shows the structure of the excitation spectrum for $N=3$ fermions discussed in detail in Ref.~\cite{Schilling21, LCLS21} to highlight the differences between the fermionic and bosonic settings.

Determining all possible lineups $l$ \eqref{lineup} is absolutely essential because each of them defines one resulting $N$-boson minimizers $\hat{\Gamma}$ in \eqref{sequence} and in that sense gives rise to one vertex of $\Sigma(\boldsymbol{w})$. According to the GOK variational principle \eqref{GOK}, the density operator corresponding to \eqref{lineup} reads
\begin{equation}
\hat{\Gamma} = \sum_{j=1}^r w_j\ket{\bd i_j}\!\bra{\bd i_j}\,.
\end{equation}
Using the spectrum of the corresponding 1RDM $\hat{\gamma}$ in \eqref{sequence} we eventually obtain the $d$-dimensional vertex
\begin{equation}\label{gen_vertex}
\bd{v}^{(l)} = \sum_{j=1}^r w_j \bd n_{\bd{i}_j}\,,
\end{equation}
where $\bd n_{\bd{i}_j} \equiv \mathrm{spec}\left(N\Tr_{N-1}\left[\ket{\bd i_j}\!\bra{\bd i_j}\right]\right)$ is an occupation number vector corresponding to the configuration $\bd i_j$. The $k$-th entry of $\bd n_{\bd{i}_j}$ equals the number of $k$'s contained in $\bd{i}_j$. Thus, by referring to the partial ordering \eqref{partial}, e.g., the lowest configuration $(1, 1, \ldots, 1)$ corresponds to the occupation number vector $\bd n_{(1, 1, \ldots, 1)} = (N, 0, \ldots, 0)$.

Finally, the spectral polytope $\Sigma(\bd w)$ follows as the convex hull of all possible permutations of the entries of all vertices $\bd v^{(l)}$ (recall \eqref{SdD}),
\begin{equation}\label{Sigma_w}
\Sigma(\boldsymbol{w}) = \mathrm{conv}\Big(\big\{\pi(\bd{v}^{(l)})\,\big\vert\,l=1, \ldots, R, \pi\in \mathcal{S}^d\big\}\Big)\,,
\end{equation}
where $\mathcal{S}^d$ denotes the permutation group of a set of $d$ elements (here the entries $\bd{v}^{(l)}$) and $R$ the number of independent  vertices $\bd v^{(l)}$. Hence, $\Sigma(\bd w)$ takes indeed the form of a polytope as anticipated above.

\begin{figure}[htb]
\includegraphics[width=.49\linewidth]{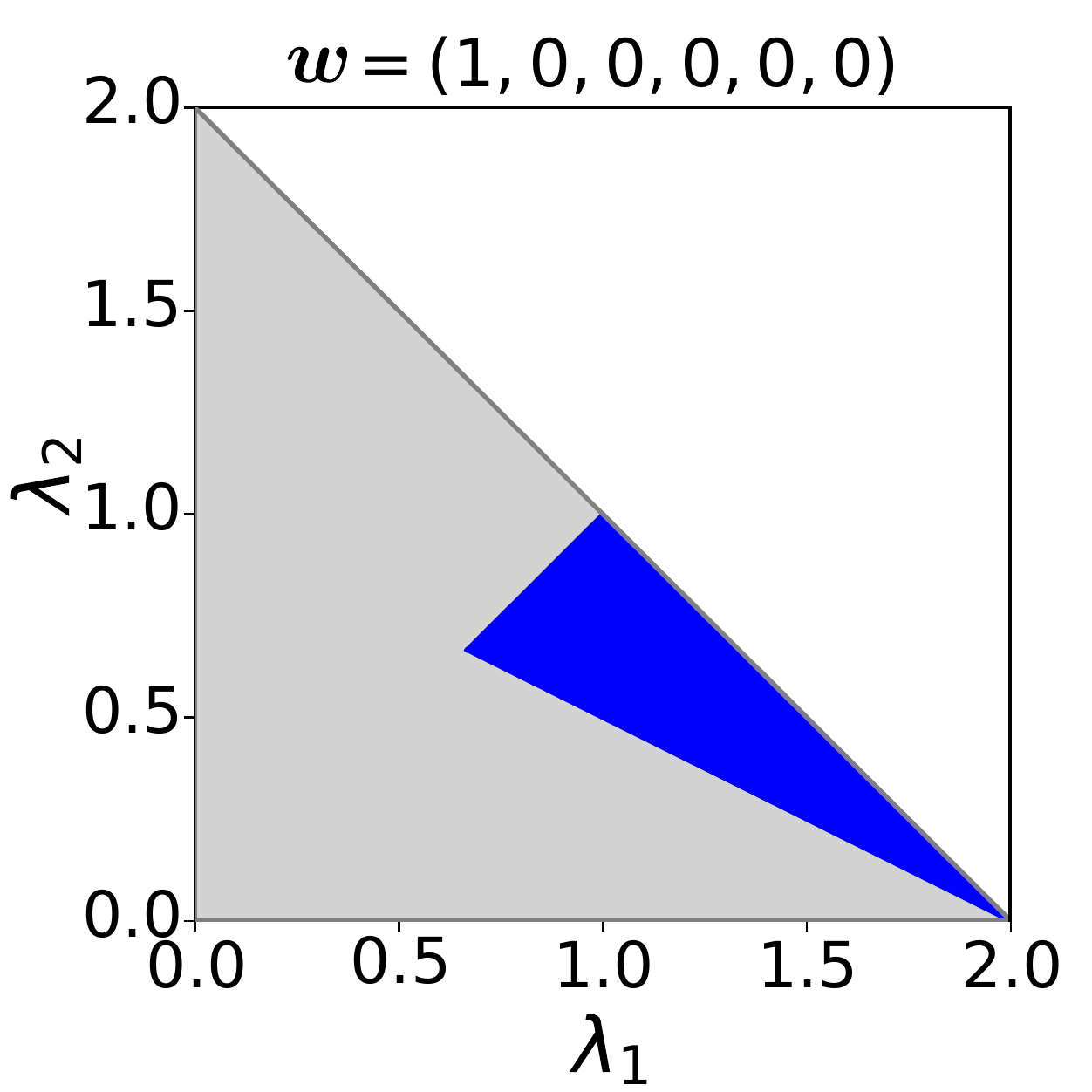}
\includegraphics[width=.49\linewidth]{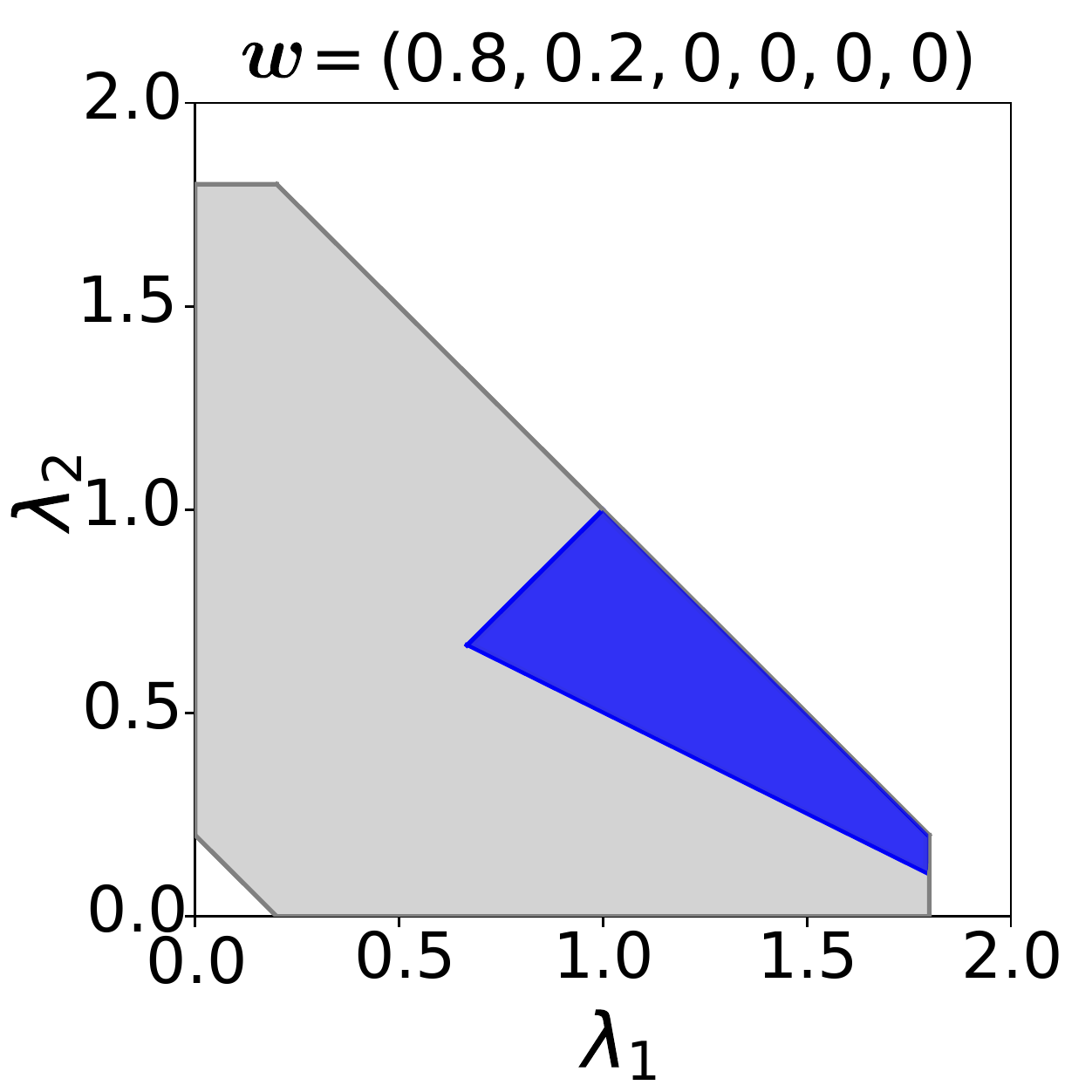}
\includegraphics[width=.49\linewidth]{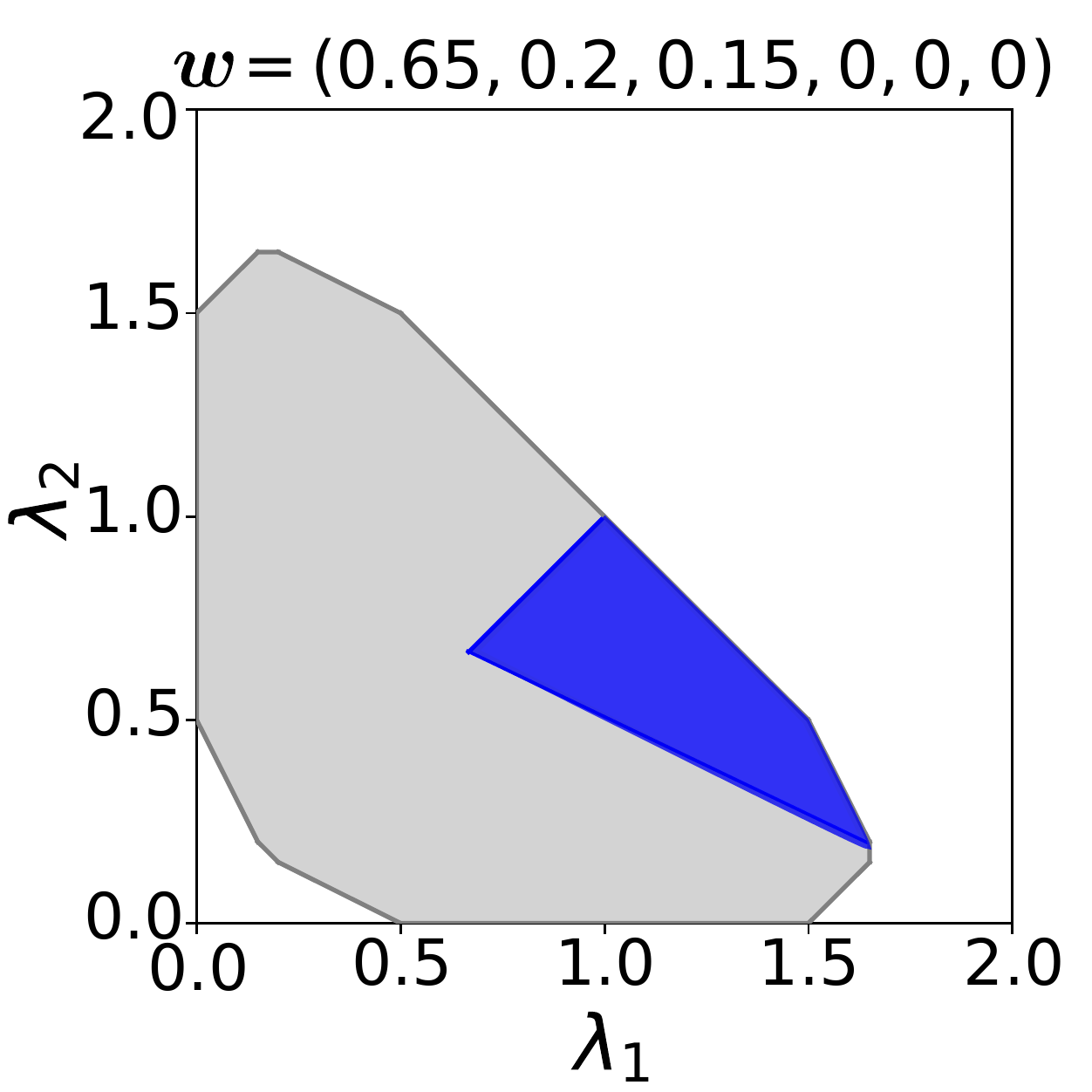}
\includegraphics[width=.49\linewidth]{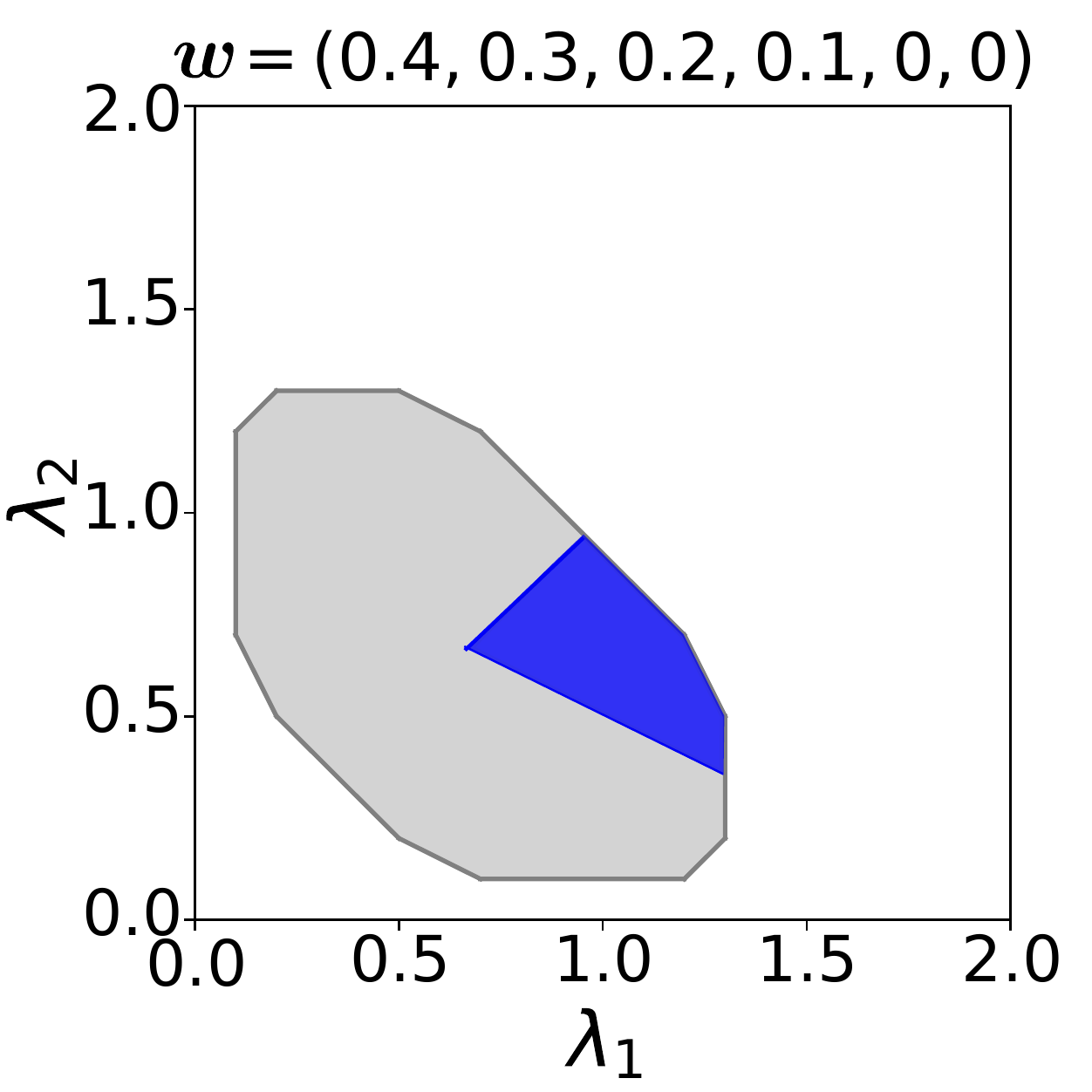}
\includegraphics[width=.49\linewidth]{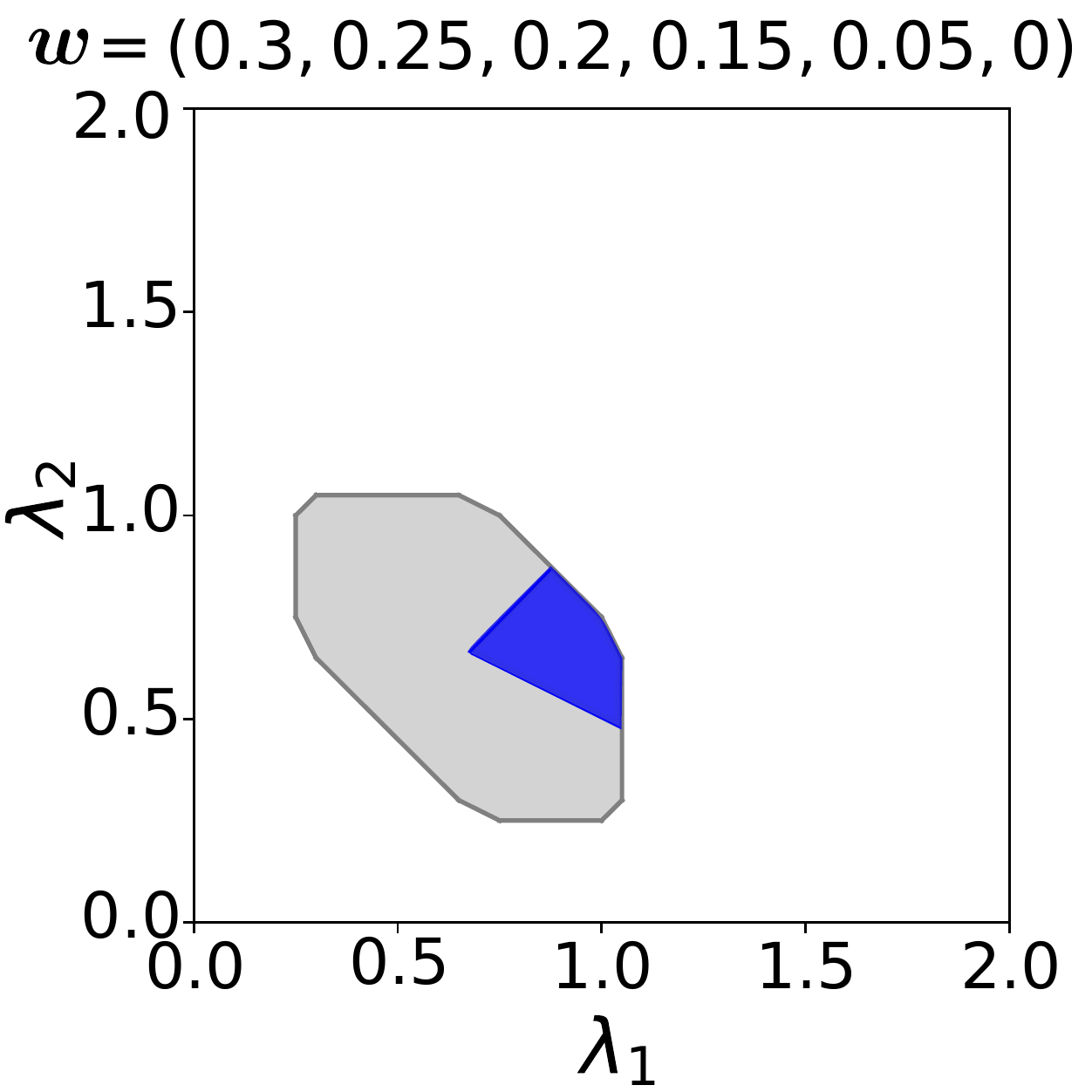}
\includegraphics[width=.49\linewidth]{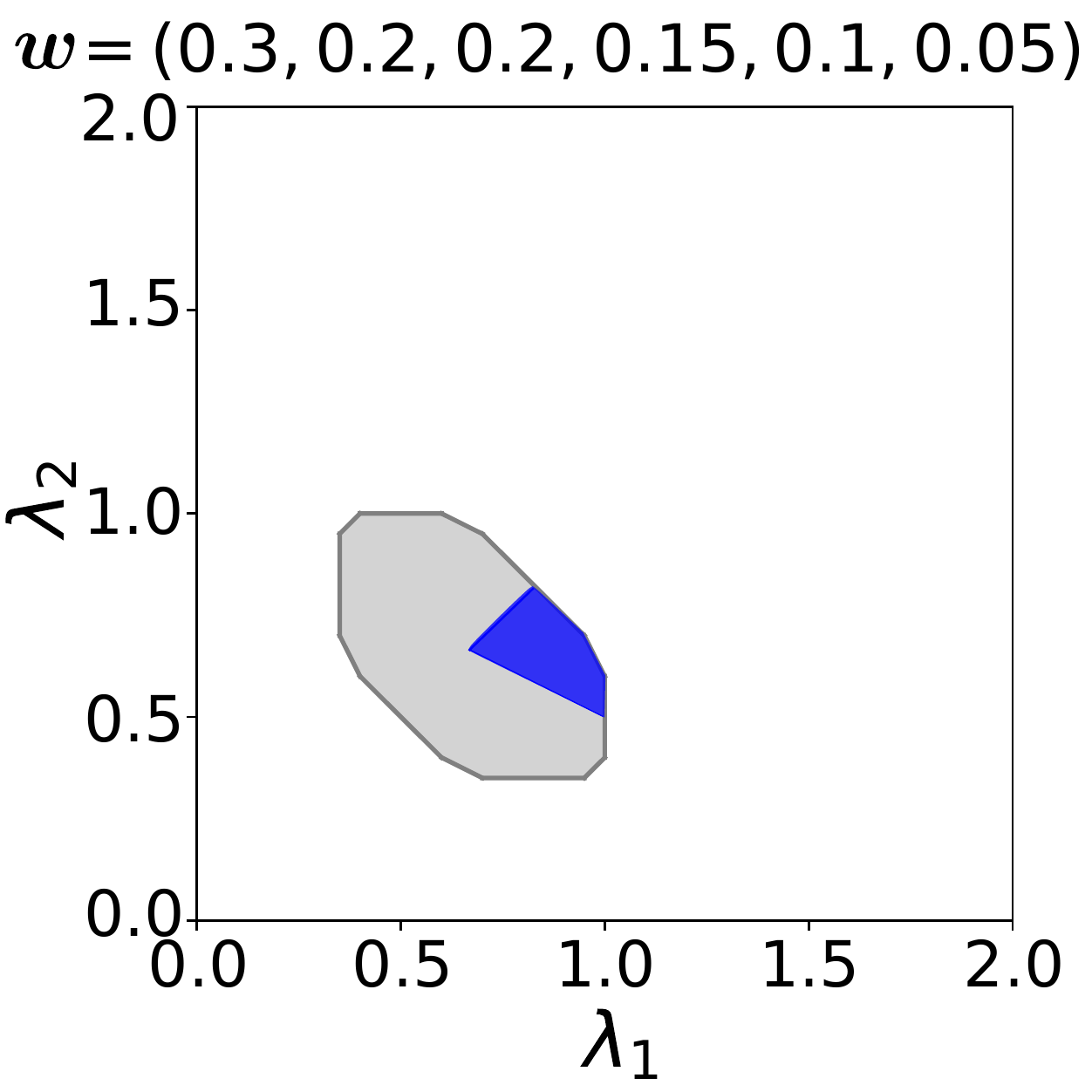}
\caption{For $(N,d) = (2, 3)$ we illustrate the inclusion relation \eqref{inclusion} of the spectral polytopes $\Sigma(\bd w)$ (grey and blue) and $\Sigma^\downarrow(\bd w)$ (blue) for several weight vectors $\bd w$.
\label{fig:inclusion}}
\end{figure}

In Fig.~\ref{fig:inclusion}, as a further illustration of the spectral polytopes, we demonstrate the important inclusion relation  Eq.~\eqref{inclusion} for the setting $(N,d)=(2,3)$. Indeed the six spectral polytopes $\Sigma(\bd{w})$ (and $\Sigma^\downarrow(\bd{w})$) are contained in each other since the six respective weight vectors $\bd{w}$ are related by majorization.

\begin{figure}[htb]
\includegraphics[width=.7\linewidth]{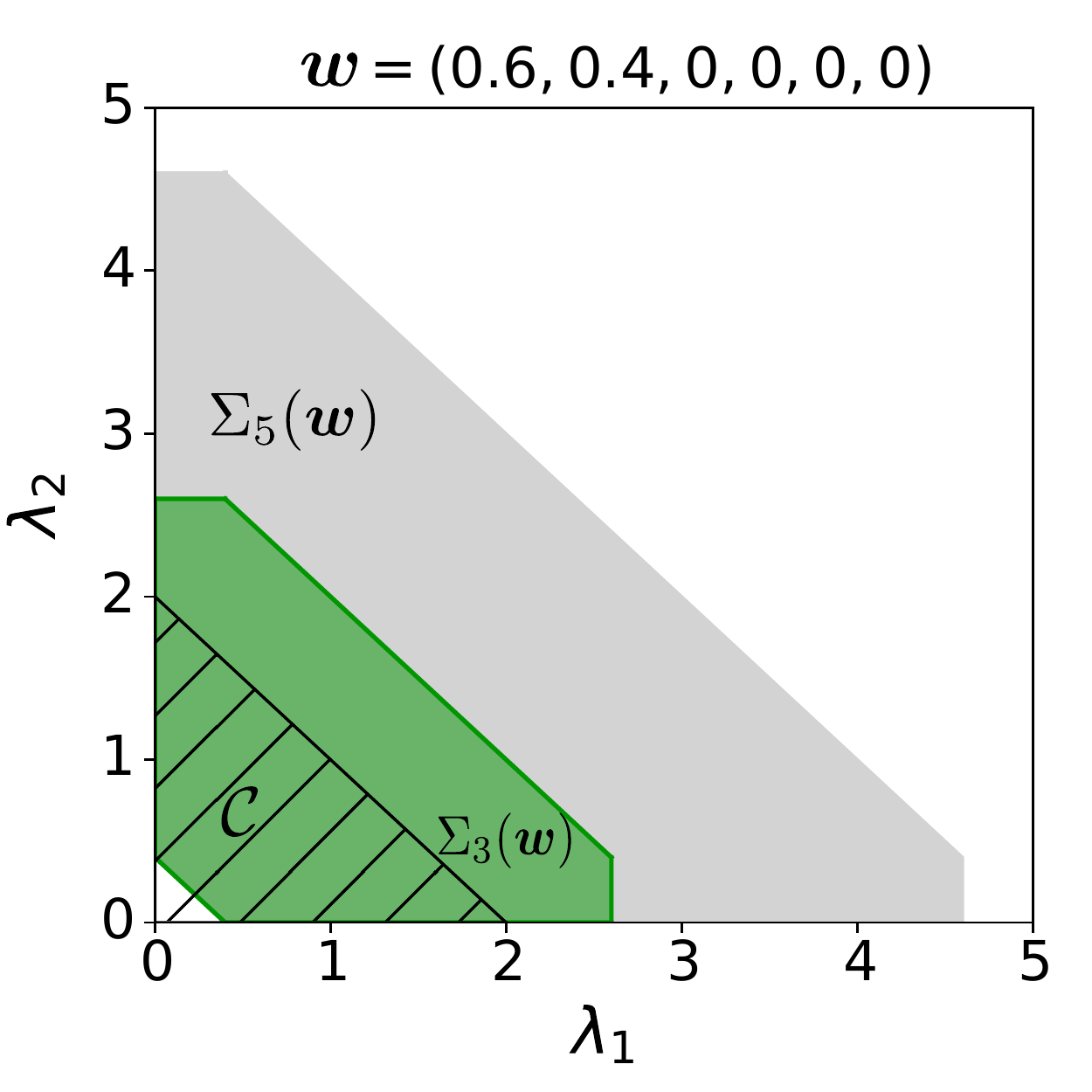}
\caption{The spectral polytope $\Sigma_5(\bd w)$ (grey) for $(N, d) = (5, 3)$ follows from the Minkowski sum of the set $\mathcal{C}$ (dashed) and the spectral polytope $\Sigma_3(\bd w)$ (green) for $(N, d)=(3, 3)$. The green polytope is contained in the grey one since $\bd{0} \in \mathcal{C}$. \label{fig:sigmaN3N5}}
\end{figure}

So far, we developed a general strategy for determining the vertex representation of the spectral polytope $\Sigma(\bd w)$ for a given total particle number $N$ and a $d$-dimensional one-particle Hilbert space $\mathcal{H}_1$.
Since experiments and theoretical studies are often performed for different particle numbers, we next comment on the relation between the spectral polytopes for different settings $(N,d)$. For $N\geq r-1$, increasing the particle number to $N^\prime > N$ just means to add to each configuration $\bd{i}_j$ in the lineups \eqref{lineup} another $N^\prime-N$ bosons in the lowest orbital $\ket{1}$. Consequently, this does not change the number of lineups and their structure. Instead, it only alters the first entry of the corresponding natural occupation number vectors $\bd{v}^{(l)}$ which indeed depends explicitly on $N$. This also implies that for fixed $d\geq r$ the vertices $\bd{v}^{(l)}$ for $N$ and $N^\prime > N$ are related through
\begin{equation} \label{vNNprime}
\bd{v}^{(l)}_{N^\prime}= \bd{v}^{(l)}_N +\delta\bd{e}_1\,,
\end{equation}
where $\delta=N^\prime-N$ and $\bd e_1 =(1, 0, \ldots)$.
Note that the polytope of a higher dimensional setting $(N^\prime,d^\prime)$ with $d^\prime > d$ can be obtained from $\Sigma_N(\bd{w})$ by the following two steps. First, one extends the $d$-dimensional vector $\bd{v}^{(l)}_N$ to a $d^\prime$-dimensional vector by adding zero entries and afterwards uses the relation \eqref{vNNprime} to increase the particle number. Yet, for the sake of simplicity we compare in the following   spectral polytopes $\Sigma_N(\bd w)$ and $\Sigma_{N^\prime}(\bd w)$ only for fixed $d$.
By applying a generalization of Rado's theorem \cite{LCLS21} (see also Eq.~\eqref{eq:genRado}) and using Eq.~\eqref{vNNprime}, we find that $\Sigma_{N^\prime}(\bd w)$ and $\Sigma_N(\bd w)$ are related through (see Appendix \ref{app:NNprime})
\begin{equation}\label{SigmaNNprimeC}
\Sigma_{N^\prime}(\bd w) \equiv \Sigma_{N}(\bd w) + \mathcal{C}\,,
\end{equation}
where $\mathcal{C}$ is a rescaled simplex with edge length $\delta$,
\begin{equation}
\mathcal{C} \equiv \mathrm{conv}(\{\pi(\delta\bd{e}_1)\,|\,\pi\in \mathcal{S}^d\})\,.
\end{equation}
The sum of the two sets in Eq.~\eqref{SigmaNNprimeC} is nothing else than the Minkowski sum of two permutation invariant polytopes which means that $\Sigma_{N^\prime}(\bd w) = \Sigma_{N}(\bd w) + \mathcal{C} = \{\bd{\lambda} + \bd{c}\,|\,\bd{\lambda}\in\Sigma_N(\bd w), \bd{c}\in \mathcal{C}\}$. Clearly, the sum of two convex sets is also convex. Also, every $\bd{\mu}\in \Sigma_{N^\prime}(\bd w)$ is correctly normalized since all $\bd{\lambda}\in \Sigma_N(\bd w)$ are normalized to $N$ and all $\bd{c}\in \mathcal{C}$ are normalized to $\delta$. In Fig.~\ref{fig:sigmaN3N5} we illustrate the Minkowski sum in Eq.~\eqref{SigmaNNprimeC} for $d=3$ and the two particle numbers $N^\prime=5$ and $N=3$. The spectral polytope $\Sigma_5(\bd w)$ (grey) is obtained by adding the elements of $\Sigma_3(\bd w)$ (green) and the rescaled simplex $\mathcal{C}$ (dashed). Due to the normalization of all occupation number vectors in $\Sigma_5(\bd w)$ and $\Sigma_3(\bd w)$ to the respective total particle number $N^\prime$ or $N$, we can omit the value of $\lambda_3$.

\section{Bosonic exclusion principle\label{sec:exclusion}}
For practical purposes, using the vertex representation of the spectral polytope $\Sigma(\bd w)$ to check whether a natural occupation number vector $\boldsymbol{\lambda}$ belongs to $\Sigma(\bd w)$ or not is highly inefficient. Yet, every polytope can equivalently be described through its halfspace representation \cite{rockafellar2015}. Then, testing the membership of $\bd \lambda$ reduces to checking only finitely many linear conditions $D_k(\boldsymbol{\lambda}^\downarrow)\geq 0$. A simple procedure for translating the vertex representation of $\Sigma(\bd w)$ into a halfspace representation is presented for $r\leq 3$ in Sec.~\ref{sec:examples}, while a more general mathematical procedure applicable to \emph{arbitrary} $r$ can be found in Ref.~\cite{CLLS21}.
These anticipated linear constraints represent a bosonic analogue of the Pauli exclusion principles since they restrict the way bosons can multiply occupy orbitals. Their existence might be surprising at first sight. This is due to the fact that the solution of the pure state $N$-representability problem for bosons is trivial, i.e.~there are no constraints on occupation numbers beyond normalization and positivity \cite{GR19,Benavides20}. This also relates to the fact that the complete hierarchy of Pauli exclusion principles \cite{LCLS21} has its origin in both the exchange symmetry and the enforced mixedness of the $N$-particle state while the bosonic constraints emerge primarily from the latter.
Furthermore, it is worth emphasizing here that the scope of these bosonic exclusion principle constraints is not restricted to the natural orbital basis. According to the Schur-Horn theorem \cite{Schur1923, Horn54} they namely apply to the occupation numbers of \emph{any} orthonormal basis. As it is explained in Sec.~\ref{sec:DFT}), they are thus potentially relevant in GOK-DFT applied to bosonic lattice models.

There are two particularly noteworthy additional structural aspects concerning the bosonic exclusion principle. First, the linear inequalities $D_k(\boldsymbol{\lambda}^\downarrow)\geq 0$ are effectively independent of the total boson number $N$ and the dimension $d$ of the one-particle Hilbert space. The only requirement for this statement to be valid is that $N$ and $d$ are large enough such that
\begin{equation}\label{Ndr}
N\geq r-1\,,\quad d\geq r\,,
\end{equation}
which can generally be assumed. This independence of the inequalities of $N$ and $d$ is in striking contrast to the generalized Pauli constraints for fermions in ground state RDMFT for pure states \cite{Schilling2018}. There, different $N$ and $d$ give rise to different inequalities which in turn makes their derivation tremendously complicated \cite{KL06, AK08, BD72, S18atoms, Kly} and their application practically impossible.
Second, there is even a hierarchy of bosonic exclusion principle constraints. To be more specific, all constraints derived for a value
$r$ are still contained in the minimal hyperplane representation for any $r^\prime>r$, i.e., they remain facet-defining. Or in other words, the constraints for $r^\prime$ are given by those for $r^\prime-1$, complemented by a few additional new ones.
A concrete illustration of this hierarchy is presented in Sec.~\ref{sec:examples} for small $r$. To demonstrate the above statement, we provide in Tab.~\ref{tab:Noineq} the number of vertices $\bd{v}^{(j)}$ and the corresponding number of inequalities for $r\leq12$. The comprehensive derivation of those halfspace conditions from the vertex representation of $\Sigma(\bd w)$ for arbitrary $r$ is presented in Ref.~\cite{CLLS21}.

\begin{table}[htb]
\centering
\begin{tabular}{ |c|c|c|c|c|c|c|c|c|c|c|c|c|c|}
\hline
$r$ & $1$  & $2$ &$3$ & $4$ & $5$&$6$ &$7$ &$8$ &$9$ &$10$&$11$ &$12$ \\
\hline
$\# \bd{v}^{(j)}$ & $1$  & $1$ &$2$ & $4$ & $8$&$17$ &$37$ &$82$ &$184$ &$418$&$967$ &$2278$ \\
\hline
$\# \text{ineq.}$ & $1$  & $2$ &$3$ & $5$ & $8$&$13$ &$22$ &$36$ &$59$ &$99$&$171$ &$299$ \\
\hline
\end{tabular}
\caption{Number of generating vertices $\bd{v}^{(j)}$ and number of exclusion principle constraints defining $\Sigma(\bd w)$ for $r\leq 12$. These numbers are independent of $N$ and $d$, provided \eqref{Ndr} is respected.}
\label{tab:Noineq}
\end{table}

\section{Derivation of inequalities for small numbers of non-vanishing weights\label{sec:examples}}
In this section, also as an illustration of Secs.~\ref{sec:characterization}, \ref{sec:exclusion},  we derive for $r\leq 3$ the bosonic exclusion principle constraints. This means to first determine the vertex representation of $\Sigma(\bd w)$ and then to turn it into a halfspace representation by using the elegant concept of vector majorization. We also explain how possible redundant inequalities can be identified in order to obtain a minimal halfspace representation of $\Sigma(\bd w)$. Throughout this section, we consider arbitrary $N$ and $d$. This then also demonstrates how the resulting facet-defining inequalities can be extended to settings of larger $N'$ and $d'$, including the important complete basis set limit $d'\rightarrow \infty$. 

\subsection{$r=1$ non-vanishing weight}
We start by deriving the hyperplane representation of the spectral polytope $\Sigma(\bd w)$ for $r=1$ and verify as a consistency check that we recover the trivial solution of the bosonic one-body pure $N$-representability problem.
As discussed in the context of Fig.~\ref{fig:spectrum}, there is only one lineup \eqref{lineup} and it consists of the single configuration $(1, \ldots, 1)$. According to \eqref{sequence}, this yields one $\bd w$-minimizer,
\begin{equation}
\hat{\Gamma}=\ket{1, \ldots, 1}\!\bra{1, \ldots,1}
\end{equation}
leading through \eqref{sequence} to the single vertex $\bd{v}=(N, 0, \ldots)$. We have thus obtained the vertex representation of $\Sigma(\bd w)$ introduced in Eq.~\eqref{Sigma_w}. Hence, $\Sigma(\bd w)$ is the permutohedron
\begin{equation}\label{permutohedron}
\Sigma(\bd w)= \mathrm{conv}(\{\pi(\bd{v})\,|\,\pi\in \mathcal{S}^d\})\,,
\end{equation}
which is nothing else than the convex hull of all possible permutations of the entries of the vertex $\bd{v}$. By Rado's theorem \cite{R52}, \eqref{permutohedron} is equivalent to
\begin{equation}\label{SRado}
\Sigma(\boldsymbol{w})= \{\boldsymbol{\lambda}\in\RR^d\,|\,\bd{\lambda}\prec \bd{v}\}\,.
\end{equation}
In particular, we have
\begin{equation}
\ebw = \{\hat{\gamma}\in \mathcal{E}^1_N\,|\,\mathrm{spec}(\hat{\gamma})\prec \bd{v}\}\,.
\end{equation}
It is exactly the majorization condition $\bd\lambda\prec \bd v$ which defines the sought-after hyperplane representation of $\Sigma(\bd w)$, i.e., the bosonic exclusion constraints. Here, $\bd\lambda\prec \bd v$ implies that
\begin{eqnarray}\label{ineq_r1}
\lambda_1^\downarrow &\leq & N\,,\nonumber \\
\lambda_1^\downarrow + \lambda_2^\downarrow &\leq & N\,,\nonumber \\
&\vdots &\nonumber\\
\sum_{i=1}^d \lambda_i^\downarrow &= &N\,.
\end{eqnarray}
Assuming $\lambda_i^\downarrow \geq 0$, all inequalities \eqref{ineq_r1} are satisfied as a direct consequence of the normalization and are therefore redundant. Thus, we indeed recover the solution of the one-body pure $N$-representability problem with the well-known trivial constraints $0\leq \lambda_i\leq N$.

\subsection{$r=2$ non-vanishing weights\label{sec:r2}}
Next, we consider $r=2$ non-vanishing weights with $w_1 + w_2 = 1$. By referring to the excitation spectrum in Fig.~\ref{fig:spectrum}, we have only one lineup \eqref{lineup} which reads
\begin{equation}
(\underbrace{1, \ldots, 1}_{N})\to (\underbrace{1,\ldots, 1}_{N-1}, 2)\,.
\end{equation}
The corresponding minimizer $\hat{\Gamma}$ is given by
\begin{equation}
\hat{\Gamma} = w_1\ket{1, \ldots, 1}\!\bra{1, \ldots, 1} + w_2\ket{1, \ldots, 1, 2}\!\bra{1, \ldots, 1, 2}
\end{equation}
and according to \eqref{sequence} and \eqref{gen_vertex} we obtain the vertex
\begin{equation}\label{v_r2}
\bd v = (N-1+w_1, w_2, 0, \ldots)\,.
\end{equation}
Since there exists only one $\bd v$ as for $r=1$, we can again apply Rado's theorem \cite{R52} to obtain a minimal hyperplane representation of $\Sigma(\bd w)$ and it follows that
\begin{equation}
\ebw = \{\hat{\gamma}\in \mathcal{E}^1_N\,|\,\mathrm{spec}(\hat{\gamma})\prec \bd{v}\}\,.
\end{equation}
Then, the majorization condition $\bd \lambda\equiv \mathrm{spec}(\hat{\gamma})\prec \bd v$ enforces the constraints
\begin{eqnarray}
\lambda_1^\downarrow &\leq & N-1+w_1\,,\nonumber \\
\lambda_1^\downarrow + \lambda_2^\downarrow &\leq & N\,,\nonumber\\
&\vdots &\nonumber \\
\sum_{i=1}^{d} \lambda_i^\downarrow&= & N\,.
\end{eqnarray}
Hence, all constraints except the first and last one are redundant, and we arrive at the minimal hyperplane representation
\begin{eqnarray}\label{consr=2}
\lambda_1^\downarrow &\leq & N-1+w_1\,,\nonumber \\
\sum_{i=1}^{d} \lambda_i^\downarrow&= & N\,.
\end{eqnarray}
Moreover, in agreement with Tab.~\ref{tab:Noineq} there is only one additional constraint in the first line of Eq.~\eqref{consr=2} compared to $r=1$. In particular, the inequalities of a minimal hyperplane representation for $r=1$ and $r=2$ represent the first two levels of the hierarchy of bosonic exclusion principles.

\subsection{$r=3$ non-vanishing weights\label{sec:r3}}
According to the excitation spectrum in Fig.~\ref{fig:spectrum}, there exist two lineups
\begin{eqnarray}
&(1)&:\,\, (\underbrace{1, \ldots, 1}_{N})\to (\underbrace{1,\ldots, 1}_{N-1}, 2)\to (\underbrace{1, \ldots, 1}_{N-1}, 3)\,,\nonumber\\
&(2)&: \,\, (\underbrace{1, \ldots, 1}_{N})\to (\underbrace{1,\ldots, 1}_{N-1}, 2)\to (\underbrace{1, \ldots, 1}_{N-2},2, 2)\,.
\end{eqnarray}
These two lineups correspond to the two vertices (recall that $w_1+w_2+w_3=1$)
\begin{eqnarray}
\bd{v}^{(1)} &=& (N-1+w_1, w_2, 1-w_1-w_2, 0, \ldots)\,,\nonumber\\
\bd{v}^{(2)} &=& (N-2 + 2w_1+w_2, 2-2w_1-w_2, 0, \ldots)\,.
\end{eqnarray}
Since there are now two vertices $\bd{v}^{(1)}$ and $\bd{v}^{(2)}$, Rado's theorem used for $r=1,2$ to obtain the hyperplane representation of $\Sigma(\bd w)$ in \eqref{Sigma_w} does not apply anymore. Instead, we can make use of a generalization of Rado's theorem introduced in Ref.~\cite{LCLS21}. It states that for vectors $\bd v^{(1)}, \ldots, \bd v^{(R)}\in \RR^d$, the polytope
\begin{equation}
\mathcal{P} = \mathrm{conv}\left(\left\{\pi(\bd v^{(j)})\,\Big\vert\,j=1, \ldots, R, \pi\in \mathcal{S}^d\right\}\right)
\end{equation}
is equivalent to
\begin{equation} \label{eq:genRado}
\mathcal{P} = \Big\{\boldsymbol{\lambda}\,\Big\vert\, \exists\,\,\text{conv. comb. }\sum_{j=1}^Rp_j\bd{v}^{(j)}\equiv \bd{v}: \boldsymbol{\lambda}\prec \bd{v}\Big\}\,.
\end{equation}
Applied to the spectral polytope $\Sigma(\bd w)$ \eqref{Sigma_w} with $R=2$, this implies that $\bd\lambda\in\Sigma(\bd w)$ if and only if there exists a convex combination $\bd{u}= q\bd{v}^{(1)} + (1-q)\bd{v}^{(2)}$ such that $\bd\lambda\prec \bd{u}$. This majorization condition leads to the constraints
\begin{eqnarray}
\lambda_1^\downarrow &\leq & N-2 + 2w_1 + w_2 + q(1-w_1-w_2)\,,\nonumber\\
\lambda_1^\downarrow + \lambda_2^\downarrow &\leq & N - q(1-w_1-w_2)\,,\nonumber\\
\lambda_1^\downarrow + \lambda_2^\downarrow + \lambda_3^\downarrow &\leq & N\,,\nonumber\\
&\vdots&\nonumber\\
\sum_{i=1}^d\lambda_i^\downarrow &= & N\,.
\end{eqnarray}
Here, all inequalities except the first two are redundant since they are automatically satisfied due to the normalization of $\bd \lambda$. To derive a minimal hyperplane representation, we need to eliminate the parameter $q$.
Therefore, we first notice that the upper bound on $\lambda_1^\downarrow$ can vary between $N-2 + 2 w_1 +  w_2$ and $N-1+ w_1$ since $q$ is restricted to $q\in[0,1]$. Moreover, the upper bound on $\lambda_1^\downarrow + \lambda_2^\downarrow$ can vary between $N$ and $N-1 +  w_1 +  w_2$.
Thus, it must always hold that $\lambda_1^\downarrow\leq N-1+ w_1$ which requires to adjust the value of $q$ according to
\begin{equation}
q\geq \frac{\lambda_1^\downarrow-N+2-2 w_1- w_2}{1- w_1- w_2}\,.
\end{equation}
This inequality is then used to tighten the upper bound on $\lambda_1^\downarrow + \lambda_2^\downarrow$ to $2(N-1) + 2w_1 + w_2 - \lambda_1^\downarrow$. Thus, our linear constraints determining a minimal hyperplane representation of the spectral polytope are given by
\begin{eqnarray}
\lambda_1^\downarrow &\leq & N-1 + w_1\,,\nonumber \\
2\lambda_1^\downarrow + \lambda_2^\downarrow &\leq & 2(N-1) + 2w_1 + w_2\,,\nonumber \\
\sum_{i=1}^d\lambda_i^\downarrow &= & N\,.
\end{eqnarray}
Thus, the second inequality is the only additional new one compared to $r=2$. This illustrates again the \textit{hierarchy} of exclusion principle constraints for bosons introduced and explained in Sec.~\ref{sec:exclusion}.

The mathematical formalism to derive systematically the halfspace representation of the spectral polytopes for larger values of $r$ is presented in Ref.~\cite{CLLS21}.

\section{DFT for excitations in lattice boson systems\label{sec:DFT}}
In this section we explain how and why the bosonic exclusion principle constraints discussed in Sec.~\ref{sec:exclusion} apply also to bosonic lattice for excitations (GOK-DFT). Let us start by recalling from a general perspective the relation and main difference between RDMFT and DFT: Compared to RDMFT, DFT restricts the class \eqref{H} of Hamiltonians of interest further to $\hat{H}(\hat{v}) \equiv \hat{v}+ \hat{t}+\hat{W}$ by fixing also the kinetic energy, i.e., only the external potential $\hat{v}$ is variable. Thus, the resulting universal functional $\Gbw$ in DFT depends on both the fixed kinetic energy $\hat{t}$ and fixed interaction $\hat{W}$ and it is universal in the sense that it is independent of $\hat{v}$. For discrete lattice systems, $\Gbw$ is a functional of the vector $\bd n \equiv(n_i)$ of the lattice site occupancies $n_i$, $\Gbw\equiv \Gbw(\bd n)$. Then, the domain of $\Gbw(\bd n)$ is the set of all vectors $\bd n$ which follow from a 1RDM $\gh$ which is (relaxed) $\wb$-ensemble $N$-representable. Since the elements of the vector $\bd n$ are nothing else than the diagonal elements of the 1RDM $\gh$ in the lattice site basis $\{\ket{i}\}$, all concepts presented in Sec.~\ref{sec:wRDMFT} can easily be translated to define the universal $\wb$-functional $\Gbw(\bd n)$ and determine its domain. According to a fundamental theorem by Schur and Horn \cite{Schur1923, Horn54}, the vector of diagonal elements of a matrix is majorized by the vector of its eigenvalues. Hence, the occupation number vector $\bd n$ is majorized by the natural occupation number vector $\bd\lambda$, i.e.
\begin{equation}\label{nversusNON}
\bd n\equiv \big(\!\langle i |\gh| i\rangle\!\big)\prec\bd\lambda \equiv \mbox{spec}(\gh)\,.
\end{equation}
Then, using a generalization of Rado's theorem \cite{LCLS21} (see also Eq.~\eqref{eq:genRado}) it follows immediately from the transitivity of the majorization that $\bd n$ must obey the same non-trivial constraints as $\bd\lambda$. In summary, our work in particular provides the first  constructive derivation of the weight dependence of the functional's domain in $\bd w$-ensemble lattice DFT. Despite the combinatorial differences in the derivation of the bosonic exclusion principle constraints, this result on lattice DFT is not specific to bosons and also applies to the fermionic case as already emphasized in Ref.~\cite{LCLS21}.

\section{Summary and Conclusions}\label{sec:concl}
We proposed and worked out a novel method for calculating excitation energies in quantum systems of correlated bosons.
Motivated by the Penrose-Onsager criterion for BEC, the exponentially complex many-boson wave function was substituted by the simpler one-particle reduced density matrix (1RDM) $\gh$: By exploiting a variational principle for density matrices with spectrum $\wb$ \cite{Gross88_1, Gross88_2, Oliveira88} we proved a corresponding generalization of the Hohenberg-Kohn \cite{HK} and Gilbert theorem \cite{Gilbert}. In that sense, we confirmed on a more formal level for both bosons and fermions the existence of a universal  $\wb$-ensemble functional. This functional theory provides access to the energies of low-lying states and their differences --- including the important ground state gap --- through the  variation of the weight distribution $\wb$ of the excited states. Accordingly, \emph{$\wb$-ensemble RDMFT} constitutes a potentially ideal framework for describing quantum phase transitions. In particular, it will be an instructive challenge to explore the \emph{nonanalytical} structural features that the universal functional needs to possess in order to correctly describe quantum phase transitions.
Yet, to first establish $\wb$-RDMFT as a practically useful method we had to circumvent the underlying $v$-representability problem that has also hampered the development of DFT and RDMFT for ground states. In order to achieve this we resorted to the Levy-Lieb constrained search formalism \cite{LE79, L83} followed by an exact convex relaxation scheme.
The latter was absolutely vital since solving the one-body $\wb$-ensemble $N$-representability problem is impossible for realistic systems sizes \cite{KL06, AK08, Kly}. It is therefore one of the main achievements of our work (see Secs.~\ref{sec:wRDMFT-LL}-\ref{sec:examples}) to circumvent the computational complexity of the respective $N$-representability problem. Inspired by the seminal work \cite{V80} by Valone on fermionic ground state RDMFT and in analogy to fermionic $\wb$-RDMFT \cite{Schilling21,LCLS21}, we achieved this by applying an exact convex relaxation to $\wb$-RDMFT. Then, in a second step we resorted to concepts from convex analysis to derive a systematic and efficient characterization of the functional's domain $\ebw$. Eventually, this revealed a hierarchy of bosonic exclusion principle constraints parameterized by the number $r$ of finite weights $w_j$. In analogy to Pauli's famous exclusion principle for fermionic pure states, these new constraints are effectively independent of the total particle number $N$ and the dimension $d$ of the one-particle Hilbert space. Our comprehensive derivation also demonstrates that the boundary of the prescribed set of admissible 1RDMs contains the entire information about the excitation spectrum of \emph{non-interacting} bosons. It follows from perturbation theoretical arguments that this statement remains approximately true for weakly interacting systems. Moreover, the bosonic exclusion principle constraints may have some broad physical relevance beyond functional theory. For instance, they could be used to dissect the statistical uncertainty of bosonic occupation numbers according to its origin to entropic thermal contributions and the interaction between the particles. An alternative prospective application is to monitor the time-evolution of open-quantum systems with an emphasis on the effect of thermalization.

\begin{acknowledgements}
We are grateful to F.~Castillo and J.P.~Labb\'e for valuable discussions.
We acknowledge financial support from the German Research Foundation (Grant SCHI 1476/1-1) (J.L., C.S.), the Munich Center for Quantum Science and Technology (C.S.) and the International Max Planck Research School for Quantum Science and Technology (IMPRS-QST) (J.L.). The project/research is also part of the Munich Quantum Valley, which is supported by the Bavarian state government with funds from the Hightech Agenda Bayern Plus.
\end{acknowledgements}

\appendix

\section{Relating spectral polytope for different particle numbers \label{app:NNprime}}

In this section, we prove that every natural occupation number vector $\bd \mu\in \Sigma_{N^\prime}(\bd w)$ corresponding to a system with $N^\prime$ bosons can be related to a $\bd \lambda\in \Sigma_N(\bd w)$ with $N^\prime> N \geq r-1$ in a unique way.
A main tool to establish the sought-after relation between the spectral polytopes for  $N$ and $N^\prime>N$ is the following generalization of Rado's theorem \cite{LCLS21} stating that for finitely many  vectors $\bd v^{(1)}, \ldots, \bd v^{(R)}\in \RR^d$, the permutation-invariant polytope
\begin{equation}
\mathcal{P} = \mathrm{conv}\left(\left\{\pi(\bd v^{(j)})\,\Big\vert\,j=1, \ldots, R, \pi\in \mathcal{S}^d\right\}\right)
\end{equation}
is equivalent to
\begin{equation} \label{genRado}
\mathcal{P} = \Big\{\boldsymbol{\lambda}\,\Big\vert\, \exists\,\,\text{conv. comb. }\sum_{j=1}^Rp_j\bd{v}^{(j)}\equiv \bd{v}: \boldsymbol{\lambda}\prec \bd{v}\Big\}\,.
\end{equation}

Also recall that the vertices $\bd v^{(j)}$ of the two spectral polytopes $\Sigma_N(\bd w)$ and $\Sigma_{N^\prime}(\bd w)$ are related through
\begin{equation}\label{vNNprime_app}
\bd v_{N^\prime}^{(j)} = \bd v_{N}^{(j)} + \delta\bd e_1\,,
\end{equation}
where $\delta= N^\prime-N$ and $\bd e_1 = (1, 0, 0, \ldots, 0)$ is a unit vector.

In the following, we consider an arbitrary but fixed number $R$ of vertices $\bd v^{(j)}$. According to two main theorems in convex analysis by Hardy, Littlewood, and P\'{o}lya \cite{HLP53} and Birkhoff, and von Neumann \cite{B46, vN53}, every $\bd \mu \in \Sigma_{N^\prime}(\bd w)$ can be written as
\begin{equation}\label{mu_decomp}
\bd \mu = \sum_{i=1}^R\sum_{\pi\in\mathcal{S}^d}q_{i, \pi}\pi(\bd{v}_{N^\prime}^{(i)})\,,
\end{equation}
where $\sum_{i=1}^R\sum_{\pi\in\mathcal{S}^d}q_{i, \pi} =1$ and $\mathcal{S}^d$ denotes the set of all permutations $\pi$ of $d$ elements. Combining this with \eqref{genRado} and \eqref{vNNprime_app} leads to
\begin{eqnarray}\label{SNNprime}
\bd{\mu} &=& \sum_{i=1}^R\sum_{\pi\in\mathcal{S}^d}q_{i, \pi}\pi(\bd{v}_{N^\prime}^{(i)}) \nonumber\\
&=& \sum_{i=1}^R\sum_{\pi\in\mathcal{S}^d}q_{i, \pi}\pi(\bd{v}_N^{(i)} + \delta\bd{e}_1) \nonumber\\
&=& \sum_{i=1}^R\sum_{\pi\in\mathcal{S}^d}q_{i, \pi}\left(\pi(\bd{v}_N^{(i)}) +\pi(\delta\bd{e}_1)\right)\nonumber\\
&\equiv & \bd{\lambda} + \delta\sum_{i=1}^R\sum_{\pi\in\mathcal{S}^d}q_{i, \pi}\pi(\bd{e}_1)\,,
\end{eqnarray}
where $\bd \lambda \in \Sigma_N(\bd w)$. The argument in \eqref{SNNprime} can be simply reverted to derive for any $\bd \lambda\in \Sigma_N(\bd w)$ a unique corresponding $\bd \mu \in \Sigma_{N^\prime}(\bd w)$. It follows that
\begin{eqnarray}\label{relNNprime}
\bd{\mu} \in \Sigma_{N^\prime}(\bd w) &=& \Sigma_N(\bd w) + \mathcal{C}\,, \nonumber \\
&=& \{\bd{\lambda} + \bd{c}\,|\,\bd{\lambda}\in\Sigma_N(\bd w), \bd{c}\in \mathcal{C}\}
\end{eqnarray}
where the set
\begin{equation}\label{Csimplex}
\mathcal{C} \equiv \mathrm{conv}(\{\pi(\delta\bd{e}_1)\,|\,\pi\in \mathcal{S}^d\})
\end{equation}
is a rescaled simplex with edge length $\delta = N^\prime-N$. Thus, according to Eq.~\eqref{relNNprime}, $\Sigma_{N^\prime}(\bd w)$ is given by the Minkowski sum of $\Sigma_{N}(\bd w)$ and $\mathcal{C}$.

\bibliography{Refs}

\end{document}